\documentclass[aps,prb,twocolumn]{revtex4-2}
\usepackage{makeidx}
\usepackage{graphicx}
\usepackage{graphics}
\usepackage{amssymb}
\usepackage{amsmath}
\usepackage{color}
\usepackage{mathtools}
\usepackage{makeidx}
\usepackage{float}

\DeclareUnicodeCharacter{03B3}{$\gamma$}

\begin{document}

\title[Test]{Microscopic Transport Theory of Cooper-pair fluctuations in a
disordered 2D Electron System with Spin-Orbit Scatterings. }
\author{Tsofar Maniv and Vladimir Zhuravlev}
\affiliation{Schulich Faculty of Chemistry, Technion-Israel Institute of Technology, Haifa 32000, Israel}

\begin{abstract}
A microscopic theory of Cooper-pair fluctuations (CPFs) in a disordered 2D
electron system with spin-orbit scatterings under parallel magnetic field is
presented in light of the observation, at low temperatures, of large
magnetoresistance (MR) above a crossover field to superconductivity in
electron-doped SrTiO$_{3}$/LaAlO$_{3}$ interfaces. It is found that in the
zero temperature limit the conventional (diagrammatic) microscopic theory of
superconducting (SC) fluctuations yields vanishing fluctuation conductivity
just above the superconducting transition. However, further analysis of the
results of the microscopic theory reveals that due to the diminishing
stiffness of the fluctuation modes in a broad range of momentum space, the
density of the CPFs, defined consistently with the time dependent
Ginzburg-Landau approach, diverges in the zero temperature limit at any
finite magnetic field. This field-induced divergence of the CPFs density,
within restricted mesoscopic regions in real space, which is relieved by
quantum tunneling and pair breaking out of their mesoscopic enclaves,
indicates that the grand canonical ensemble underlying the microscopic
theory is unsubstantiated. A dynamical equilibrium between the condensed
CPFs in real-space mesoscopic puddles and the rarefying system of unpaired
electrons controls the residual normal-state conductivity at magnetic fields
above the SC transition. It has been, therefore, concluded that under the
diminishing fluctuation paraconductivity upon increasing magnetic field the
density of the normal-state electrons is also suppressed (due to charge
transfer to the localized CPFs) and so, resulting from electron
localization, the overall MR is strongly enhanced.
\end{abstract}

\maketitle

\section{Introduction}

It was shown recently \cite{MZPRB2021},\cite{MZJPC2023} that Cooper-pair
fluctuations (CPFs) in a 2D electron system with strong spin-orbit
scatterings can lead at low temperatures to pronounced magnetoresistance
(MR) above a crossover field to superconductivity. Employing the time
dependent Ginzburg-Landau (TDGL) functional approach the model was applied
to the high mobility electron systems formed in the electron-doped
interfaces between two insulating perovskite oxides---SrTiO$_{3}$ and LaAlO$%
_{3}$ \cite{Ohtomo04},\cite{Caviglia08}, showing good quantitative agreement
with a large body of experimental sheet-resistance data obtained under
varying gate voltage \cite{Mograbi19} (see also \cite{RoutPRL2017},\cite%
{ManivNatCommun2015}).

In the present paper we approach the same problem from the point of view of
the conventional (diagrammatic) microscopic theory of fluctuations in
superconductors, developed by Larkin and Varlamov (LV) \cite{LV05}.
Consistently with the state-of-the-art microscopic theory of superconducting
fluctuations at very low temperatures \cite{GalitLarkinPRB01}, \cite%
{Glatzetal2011}, \cite{Lopatinetal05}, \cite{Khodas2012}, our model
calculations yields corrections to the normal-state conductivity, which can
not account for the observed pronounced MR effect. However, further analysis
of the results of the microscopic theory reveals that the density of
Cooper-pair fluctuations (CPFs), defined consistently with the TDGL
approach, is a key function of field and temperature for understanding the
large MR phenomenon. It is then found that, in the zero temperature limit at
finite magnetic field, the stiffness of the fluctuation modes diminishes in
a broad range of momentum space, so that the CPFs density diverges within
restricted mesoscopic regions in real space. This divergence is relieved by
quantum tunneling of CPFs and pair breaking out of their mesoscopic enclaves 
\cite{MZPRB2021},\cite{MZJPC2023}. Under these circumstances the grand
canonical ensemble underlying the microscopic BCS-GL theory is
unsubstantiated\ and charge exchange between regions of CPFs and (unpaired)
normal-state electrons should be considered under the constraint of the
total electron number conservation.

It has been, therefore, concluded that during the extended life-time of CPFs
in their mesoscopic enclaves, when the fluctuation paraconductivity
diminishes, the density of the normal-state conduction electrons is also
suppressed (due to charge transfer to the localized CPFs) and so, due to
electron localization, yielding overall large MR. The crossover to this low
temperature state of pronounced MR is, therefore, associated with the
tendency of the nearly uniform system of CPFs at zero field to formation,
under increasing field, of highly inhomogeneous system of condensed CPFs
puddles.

The paper is organized as follows: In Sec.II we present the model of a 2D
electron system with strong spin-orbit scatterings employed in this paper.
The model is then applied in Sec.III to the microscopic (zero field) LV
theory at very low temperature. Subsequently, in the same section, the
results of the microscopic (diagrammatic) theory for the fluctuation
conductivity are compared to those of the corresponding TDGL functional
approach, revealing important connections between the two approaches. In
Sec.IV we present an extension of the microscopic theory of fluctuation
conductivity to finite magnetic field, which is further developed in Sec.V
by introducing the concept of CPFs density. The phenomenon of field-induced
condensation of CPFs at very low temperatures, revealed in analyzing this
concept, and its consequent phenomenon of crossover to localization of CPFs,
are also discussed in Sec.V. A detailed discussion of the physical
ramifications of the main findings of this paper and concluding remarks
appear in SecVI.

\begin{center}
\begin{figure*}[tbp]
\includegraphics[width=3.5in]{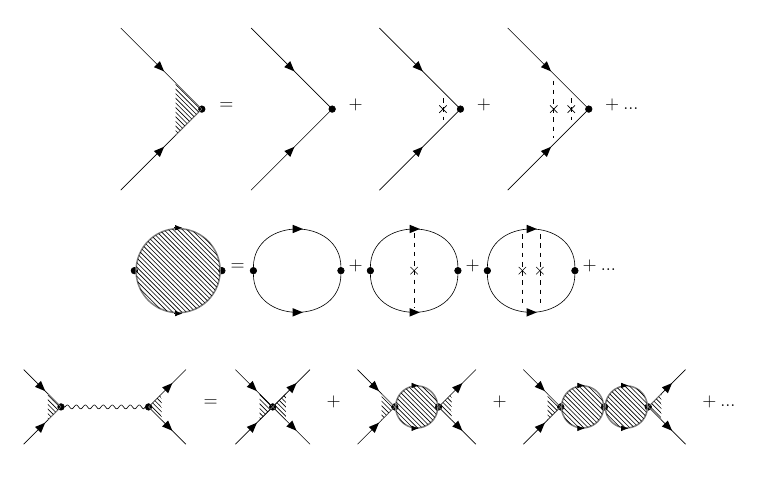} %
\includegraphics[width=2.5in]{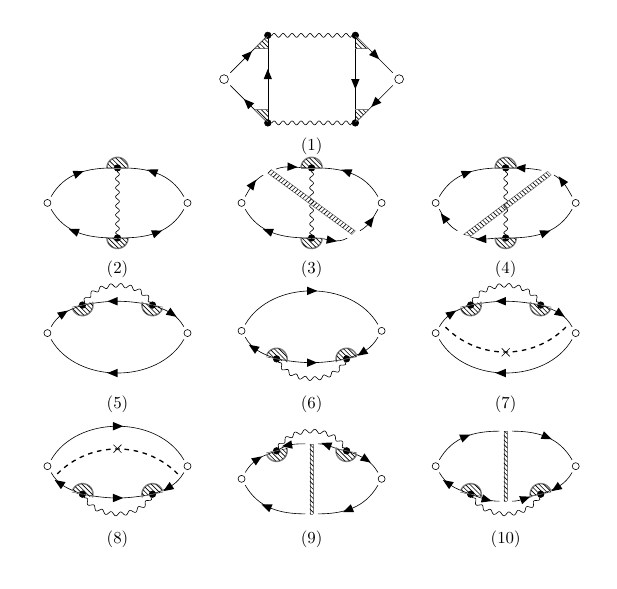}
\caption{{}Right figure: \ Leading-order Feynman diagrams of the
current-current correlator contributing to the fluctuation conductivity.
Small empty circles are bare current vertices, solid lines are impurity
average normal-state Green's functions, wavy lines are fluctuation
propagators, small full circles are bare pairing vertices, dashed areas
stand for impurity ladders between two electron lines, dashed lines with
central crosses are additional impurity renormalizations. Left figure:
External (upper row) and internal (middle row) impurity-scattering
renormalization of pairing vertices appearing in the Feynman diagrams on the
right and in the fluctuation propagator (lower row on the left).}
\label{fig1}
\end{figure*}
\end{center}

\section{The model}

The model employed in this paper, following Refs. \cite{MZPRB2021},\cite%
{MZJPC2023}, which have been motivated by the perovskite oxides electronic
interface state investigated in Ref. \cite{Mograbi19}, consists of a thin
rectangular film of disordered electron system, under a strong magnetic
field $H$, applied parallel to the conducting plane. Disorder is due to
impurity scatterings, including both nonmagnetic potential scattering and
spin-orbit scattering \cite{Abrikosov62}, \cite{Klemm75}. Superconductivity
in this system is governed by the interplay between the Zeeman spin
splitting energy,$\mu _{B}H$ and the spin-orbit scattering rate, $1/\tau
_{SO}\equiv \varepsilon _{SO}/\hbar $ (see a detailed description in early
papers dealing with similar 3D systems \cite{Maki66}, \cite{FuldeMaki70}, 
\cite{WHH66}), where the corresponding spin-flip scattering processes
effectively suppress the pair-breaking effect of the Zeeman spin-splitting.
The nonmagnetic potential scatterings influence the CPFs only through the
coherence length, i.e. via the effect of the scattering rate on the
electronic diffusion constant (see Ref.\cite{Maki66}).

The spin-orbit impurity scattering matrix employed, following Ref.\cite%
{MZPRB2021}, has been a reasonable model for the strong spin-orbit
interaction of the Ti 3d conduction electrons with lattice ions in SrTiO$%
_{3} $(\cite{CavigliaPRL2010}). For the (111) LaAlO$_{3}$/SrTiO$_{3}$
interface employed in Ref.\cite{Mograbi19} the spin-relaxation time $\tau
_{SO}$, was found (see Ref.\cite{RoutPRL2017}) to follow the (Elliott-Yafet)
relation \cite{CavigliaPRL2010},\cite{ZuticRMP2004}: $\tau _{SO}\sim \tau
_{p}$, where $\tau _{p}$ is the elastic momentum relaxation time, in a
region of gate voltages where crossovers from superconductivity to large MR
have been observed \cite{Mograbi19}.

Under these circumstances the use of $\tau _{SO}$ as a single relaxation
time in our calculations seems reasonable. However, one can readily show, on
the basis of results published long time ago \cite{Maki66}, \cite{WHH66}
(and reproduced for a 2D system in \cite{unpublished}), that our model
adequately describes also the usual situations where the non-magnetic
potential scatterings dominate the relaxation processes, the only
modification due to the non-magnetic scatterings appears through the
electronic diffusion constant controlling the fluctuation kinetic energy.

We use a reference of frame in which the conducting interface is in its $z-x$
plane, the film thickness (along the $y$ axis) is $d$, and $\mathbf{E}%
\mathbf{\mathbf{=}}\widehat{x}E$, \ $\mathbf{B}\mathbf{=}\widehat{z}H$ are
the in-plane electric and magnetic fields, respectively. The transport
calculations are carried out in the linear response approximation with
respect to the electric field and impurity scattering is treated in the
dirty limit, i.e. for $k_{B}T\tau _{SO}/\hbar \ll 1$ (see also, below, an
extension of the dirty limit condition in the presence of magnetic field).
Typical values of the parameters of the electronic system used are: $%
E_{F}\simeq 7{\mathit{meV}},\varepsilon _{SO}\simeq 3{\mathit{meV}},d\simeq
10^{-9}{\mathit{m}}$, and the characteristic magnetic field of the crossover
to superconductivity at zero temperature is $H_{c\parallel 0}=4.5{\mathit{T}}
$. Under these circumstances the Zeeman spin-splitting energy: $\mu
_{B}H_{c\parallel 0}\simeq 0.26{\mathit{meV}}$, is much larger than the
diamagnetic (kinetic) energy: $\hslash D\left( deH_{c\parallel 0}/\hbar
\right) ^{2}\simeq 8\times 10^{-3}{\mathit{meV}}$, where $D\equiv \tau
_{SO}v_{F}^{2}/2=\hbar E_{F}/\varepsilon _{SO}m^{\ast }$ is the electronic
diffusion coefficient, with the electronic band effective mass $m^{\ast }$
close to the free electron mass $m_{e}$. We, therefore neglect the
diamagnetic energy in our analysis below.

Our model is similar to the models employed in both Refs.\cite{Lopatinetal05}
and \cite{Khodas2012}, but differs in an important aspect. In our model
impurity scatterings are dominated by spin-orbit interactions whereas in
Refs.\cite{Lopatinetal05}, \cite{Khodas2012} spin-orbit interaction was
absent.

\section{TDGL functional approach vs. microscopic theory at zero field}

\subsection{The microscopic Larkin-Varlamov (diagrammatic) theory}

In this subsection we introduce the formalism, employed by LV for the zero
magnetic field situation at temperatures above the zero-field transition
temperature $T_{c0}$, in a general form which allows extension to finite
field at low temperatures well below $T_{c0}$. It is implicitly assumed,
throughout this paper, that the critical shift parameter $\varepsilon $ (or $%
\varepsilon _{H}$ for the finite-field situations, see below for more
details) includes high-order terms in the Gorkov GL expansion,
self-consistently in the interaction between fluctuations \cite{UllDor90},%
\cite{UllDor91}, as done in Ref.\cite{MZPRB2021}.

As indicated in Sec.II, the incorporation of such interactions
self-consistently into the equation determining the "critical" shift
parameter avoids it vanishing at any field and temperature and so, in
particular, removes the quantum critical point (\cite{Lopatinetal05}, \cite%
{Khodas2012}). Our motivation in selecting this approach is the absence of
genuine zero resistance in the experimental data reported in Ref.\cite%
{Mograbi19}, even at zero field.

\subsubsection{The Aslamazov-Larkin paraconductivity diagram}

The principal contribution to the conductivity due to CPFs
(paraconductivity) in the microscopic (diagrammatic) theory is associated
with the Aslamazov-Larkin (AL) diagram \cite{AL68} (Number 1 in Fig.1, see
also Fig.2). The corresponding static Aslamazov-Larkin conductivity $\sigma
_{AL}^{LV}$ is obtained from the retarded current-current correlator $%
Q_{xx}^{LV\left( 1\right) }\left( \omega \right) $ : $\sigma
_{AL}^{LV}=\lim_{\omega \rightarrow 0}\frac{i}{\omega }\left[
Q_{xx}^{LV\left( 1\right) }\left( \omega \right) -Q_{xx}^{LV\left( 1\right)
}\left( 0\right) \right] $, where $Q_{xx}^{LV\left( 1\right) }\left( \omega
\right) $ is the analytic continuation of the Matzubara imaginary
time-ordered correlator, i.e.: $Q_{xx}^{LV\left( 1\right) }\left( \omega
\right) =Q_{xx}^{LV(1)}\left( i\Omega _{\nu }\rightarrow \omega +i\delta
\right) $. The corresponding time-ordered correlator in imaginary frequency
representation can be written as \cite{LV05}: 
\begin{eqnarray}
&&Q_{xx}^{LV(1)}\left( i\Omega _{\nu }\right) =-4e^{2}k_{B}T\sum
\limits_{k=-\infty }^{\infty }\frac{1}{d}\left( \frac{1}{2\pi }\right)
^{2}\int d^{2}q  \label{Q^LV(1)} \\
&&\times \mathcal{D}\left( q,\Omega _{k}+\Omega _{\nu }\right) B_{x}\left(
q,\Omega _{k},\Omega _{\nu }\right) \mathcal{D}\left( q,\Omega _{k}\right)
B_{x}\left( q,\Omega _{k},\Omega _{\nu }\right)  \notag
\end{eqnarray}%
where: 
\begin{equation}
\mathcal{D}\left( q,\Omega _{k}\right) =\frac{1}{N_{2D}\left[ \varepsilon +%
\frac{\pi \hbar }{8k_{B}T}\left( Dq^{2}+\left \vert \Omega _{k}\right \vert
\right) \right] }  \label{D(q,Omk)}
\end{equation}%
is the fluctuation propagator in wavenumber-(Matsubara) frequency
representation, $\varepsilon =\ln \left( T/T_{c0}\right) $, $T_{c0}$ is the
zero-field transition temperature, $D=\tau _{SO}v_{F}^{2}/2$ is the
electronic diffusion coefficient, and $v_{F}=\hbar k_{F}/m^{\ast }$ is the
Fermi velocity. Here $\Omega _{k}=2kk_{B}T/\hbar ,\Omega _{\nu }=2\nu
k_{B}T/\hbar ,$ $k=0,\pm 1,\pm 2,...,$\ \ \ \ $\nu =0,1,2,....$ are bosonic
Matsubara frequencies and $N_{2D}=m^{\ast }/2\pi \hbar ^{2}$ is the
single-electron density of states (DOS), with an effective mass $m^{\ast }$.
It should be also noted that Eq.\ref{D(q,Omk)} is a small-wavenumber
approximation of the fluctuation propagator at zero field, derived in Ref.%
\cite{MZPRB2021} (see Eq.\ref{D(q)^H}).

In Eq.\ref{Q^LV(1)} the effective current vertex part (see Fig.2) is given
by: 
\begin{widetext}
\begin{eqnarray}
B_{x}\left( q,\Omega _{k},\Omega _{\nu }\right) &=&k_{B}T\sum
\limits_{n=-\infty }^{\infty }\lambda \left( q,\omega _{n}+\Omega _{\nu
},\Omega _{k}-\omega _{n}\right) \lambda \left( q,\omega _{n},\Omega
_{k}-\omega _{n}\right) \times  \label{B_x(q,om)} \\
&&\left( \frac{1}{2\pi }\right) ^{2}\int d^{2}pG\left( \mathbf{p},\omega
_{n}+\Omega _{\nu }\right) G\left( \mathbf{p},\omega _{n}\right) G\left( 
\mathbf{q-p},\Omega _{k}-\omega _{n}\right) v_{x}\left( \mathbf{p}\right) 
\notag
\end{eqnarray}%
\end{widetext}where $\omega _{n}=\left( 2n+1\right) k_{B}T/\hbar ,n=0,\pm
1,\pm 2,...$ , is a fermionic Matsubara frequency, and $\lambda $ stands for
a three-leg vertex of two electron lines and a fluctuation line,
renormalized by impurity-scattering ladder (see left Fig.1). A
single-electron line corresponds to the Green's function: $G\left( \mathbf{p}%
,\omega _{n}\right) =1/\left[ i\hbar \widetilde{\omega }_{n}-\hbar
^{2}\left( p^{2}-k_{F}^{2}\right) /2m^{\ast }\right] $, whereas the "bare"
current vertex is: $v_{x}\left( \mathbf{p}\right) =\hbar p_{x}/m^{\ast }$.

At zero magnetic field, following LV, we have: 
\begin{equation}
\lambda \left( q,\omega _{n},\Omega _{k}-\omega _{n}\right) =\frac{\left
\vert \widetilde{\omega }_{n}-\left( \widetilde{\Omega _{k}-\omega }%
_{n}\right) \right \vert }{\left \vert 2\omega _{n}-\Omega _{k}\right \vert +%
\frac{v_{F}^{2}q^{2}}{2\tau _{SO}\left \vert \widetilde{\omega }_{n}-\left( 
\widetilde{\Omega _{k}-\omega }_{n}\right) \right \vert ^{2}}}
\label{lambda}
\end{equation}%
where $\widetilde{\omega }_{n}=\omega _{n}+\left( 1/2\tau _{SO}\right)
sign\left( \omega _{n}\right) $, whereas $B_{x}\left( q,\Omega _{k};\Omega
_{\nu }\right) $ is approximated by taking: $\Omega _{\nu }=0,$ $\Omega
_{k}=0$, so that: 
\begin{eqnarray*}
&&B_{x}\left( q,\Omega_{k},\Omega _{\nu }\right) \simeq B_{x}\left(
q,0,0\right) =k_{B}T\sum \limits_{n = -\infty }^{\infty }\lambda^{2}\left(
q,\omega _{n},-\omega _{n}\right) \\
&& \left( \frac{1}{2\pi }\right) ^{2}\int d^{2}pG\left( \mathbf{p},\omega
_{n}\right) G\left( \mathbf{p},\omega _{n}\right) G\left( \mathbf{q-p}%
,-\omega _{n}\right) v_{x}\left( \mathbf{p}\right)
\end{eqnarray*}

and:

\begin{equation}
\lambda \left( q,\omega _{n},-\omega _{n}\right) =\frac{\left \vert 
\widetilde{\omega }_{n}\right \vert }{\left \vert \omega _{n}\right \vert +%
\frac{v_{F}^{2}q^{2}}{4\tau _{SO}\left \vert 2\widetilde{\omega }_{n}\right
\vert ^{2}}}\simeq \frac{\left \vert \widetilde{\omega }_{n}\right \vert }{%
\left \vert \omega _{n}\right \vert +\frac{1}{2}Dq^{2}}  \label{lambda(q)}
\end{equation}

Performing the integration over the electronic wavevector and the Matsubara
frequency summation we find: 
\begin{equation}
B_{x}\left( q,0,0\right) \approx -N_{2D}\tau _{SO}v_{F}^{2}q_{x}\frac{1}{%
4\pi k_{B}T}\psi ^{\prime }\left( \frac{1}{2}+\frac{D}{4\pi k_{B}T}%
q^{2}\right)  \label{B_x(q,0,0)}
\end{equation}
where $\psi$ is digamma function.

Neglecting the $q$ dependence, i.e. taking $B_{x}\left( q\rightarrow
0,0,0\right) \rightarrow -2N_{2D}\eta _{\left( 2\right) }q_{x}$, with:

\begin{equation}
\eta _{\left( 2\right) }=\frac{\pi \hbar D}{8k_{B}T}  \label{eta(2)}
\end{equation}%
the corresponding expression for the zero-field AL conductivity is found to
be identical to the LV integral form of the well-known result:

\begin{equation*}
\sigma _{AL}^{LV}=\frac{e^{2}}{8\hbar d}\int d\left( \eta _{\left( 2\right)
}q^{2}\right) \frac{\left( \eta _{\left( 2\right) }q^{2}\right) }{\left(
\varepsilon +\eta _{\left( 2\right) }q^{2}\right) ^{3}}
\end{equation*}

\begin{figure}[tbh]
\includegraphics[width=2in]{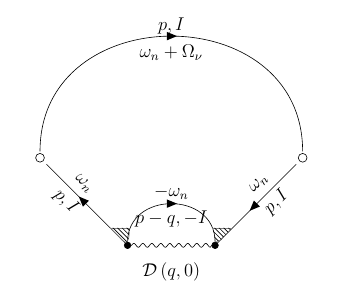} \includegraphics[width=3in]{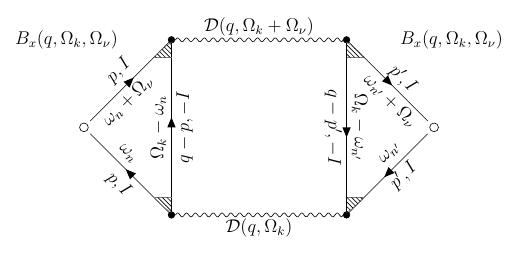}
\caption{{}Lower figure: AL diagram consisting of effective current vertices
($B_{x}$) and fluctuation propagators ($\mathcal{D}$). Upper figure: DOS
conductivity diagram illustrating the mechanism in which electron time of
coherence is suppressed due to electron scattering by background electrons
via virtual exchange of CPF with $q\neq 0$ and $I\neq 0$. }
\label{fig2}
\end{figure}

\subsubsection{The DOS conductivity diagrams}

The next leading contributions to the fluctuation conductivity in the LV
scheme corresponds to four bubble-shaped diagrams (numbers 5-8 in Fig.1),
termed density-of-states diagrams, due to the fluctuation self-energy
insertion to single-electron lines. Other, topologically distinct diagrams
of the same order in the fluctuations propagators, (see in Fig.1 the
diagrams numbered 2-4), well-known as the Maki-Thompson \cite{MakiPTP1968},%
\cite{ThompsonPRB1970}, type of diagrams, have been disregarded in our model
since the strong spin-orbit scatterings, which characterize the SrTiO$_{3}$%
/LaAlO$_{3}$ interfaces under consideration, are known to suppress their
overall contribution \cite{LV05} (See Appendix A). Furthermore, as explained
in LV, the DOS diagrams 9,10 can also be neglected, however at very low
temperature and close to the quantum critical point they dominate the
conductivity (see Refs.\cite{Lopatinetal05}, \cite{Glatzetal2011}, \cite%
{Khodas2012}). We will return to the issue of quantum critical fluctuations
in Sec.VI.

The current-current correlator corresponding, e.g. to diagram 5 (see also
Fig.2 upper part) is written as:

\begin{eqnarray}
&&Q_{xx}^{LV(5)}\left( i\Omega _{\nu }\right) =-2e^{2}k_{B}T\sum
\limits_{k=-\infty }^{\infty }\frac{1}{d}\left( \frac{1}{2\pi }\right) ^{2} 
\notag \\
&&\times \int d^{2}q\Sigma _{xx}^{(5)}\left( q,\Omega _{k},\Omega _{\nu
}\right) \mathcal{D}\left( q,\Omega _{k}\right)  \label{Q^LV(5)}
\end{eqnarray}%
where:

\begin{eqnarray}
&&\Sigma _{xx}^{(5)}\left( q,\Omega _{k},\Omega _{\nu }\right) =k_{B}T\sum
\limits_{n=-\infty }^{\infty }\lambda ^{2}\left( q,\omega _{n},\Omega
_{k}-\omega _{n}\right)  \notag \\
&&\times I_{xx}^{(5)}\left( \mathbf{q},\omega _{n},\Omega _{k},\Omega _{\nu
}\right)  \label{Sig_xx^(5)}
\end{eqnarray}%
and:

\begin{eqnarray}
&&I_{xx}^{(5)}\left( \mathbf{q},\omega _{n},\Omega _{k},\Omega _{\nu
}\right) =\left( \frac{1}{2\pi }\right) ^{2}\int d^{2}pG^{2}\left( \mathbf{p}%
,\omega _{n}\right)  \notag \\
&&G\left( \mathbf{q-p},\Omega _{k}-\omega _{n}\right) G\left( \mathbf{p}%
,\omega _{n}+\Omega _{\nu }\right) v_{x}^{2}\left( \mathbf{p}\right)
\label{Ixx^(5)}
\end{eqnarray}

Neglecting dynamical (quantum critical) fluctuations by restricting
consideration only to the single term with $\Omega _{k}=0$ in the summation
over $k$ in Eq.\ref{Q^LV(5)}, performing the integration over the electron
wavevector in Eq.\ref{Ixx^(5)} and the Matsubara frequency summation in Eq.%
\ref{Sig_xx^(5)}, we get after the analytic continuation: $\Omega _{\nu
}\rightarrow -i\omega $:

\begin{eqnarray}
&\Sigma _{xx}^{(5)}\left( q,\Omega _{k}=0,\Omega _{\nu }\rightarrow -i\omega
\right) \approx -\frac{i\omega \tau _{SO}^{2}v_{F}^{2}N_{2D}}{8\pi
k_{B}T\hbar }\times  \notag \\
&\left[ \psi ^{\prime }\left( \frac{1}{2}+\frac{\hbar }{4\pi k_{B}T\tau _{SO}%
}\right) -\frac{3\hbar }{4\pi k_{B}T\tau _{SO}}\psi ^{\prime \prime }\left( 
\frac{1}{2}+\frac{\hbar Dq^{2}}{4\pi k_{B}T}\right) \right]  \notag \\
&\approx \frac{i\omega DN_{2D}}{4\pi k_{B}T}\frac{3}{4\pi k_{B}T}\psi
^{\prime \prime }\left( \frac{1}{2}+\frac{\hbar Dq^{2}}{4\pi k_{B}T}\right)
\label{Sigma_xx^(5)}
\end{eqnarray}%
Further simplification, consistent with the procedure used in evaluating the
AL contribution, is achieved by neglecting the $q$ dependence of $\Sigma
_{xx}^{(5)}\left( q,\Omega _{k}=0,\Omega _{\nu }\rightarrow -i\omega \right) 
$ in Eq.\ref{Q^LV(5)}, that is: {\small 
\begin{equation*}
\Sigma _{xx}^{(5)}\left( q\rightarrow 0,\Omega _{k}=0,\Omega _{\nu
}\rightarrow -i\omega \right) \rightarrow \frac{i\omega DN_{2D}}{4\pi k_{B}T}%
\frac{3}{4\pi k_{B}T}\psi ^{\prime \prime }\left( \frac{1}{2}\right)
\end{equation*}%
} which yields for the corresponding contribution to the fluctuation
conductivity:{\small 
\begin{equation*}
\sigma _{xx}^{LV\left( 5\right) }=-\lim_{\omega \rightarrow 0}\frac{%
Q_{xx}^{LV\left( 5\right) }\left( \omega \right) }{i\omega }\simeq -\frac{%
3e^{2}}{4\hbar d}\frac{14\zeta \left( 3\right) }{\pi ^{4}}\int \frac{d\left(
\eta _{\left( 2\right) }q^{2}\right) }{\varepsilon +\left( \eta _{\left(
2\right) }q^{2}\right) }
\end{equation*}%
} where we have used the identity: $\psi ^{\prime \prime }\left( 1/2\right)
=-\sum \limits_{n=0}^{\infty }1/\left( n+1/2\right) ^{3}=-14\zeta \left(
3\right) $.

An identical result is obtained for diagram No.6 (see Fig.1), i.e.: $\sigma
_{xx}^{LV\left( 6\right) }=\sigma _{xx}^{LV\left( 5\right) }$. \ A similar
method of calculation yields for the 7-th diagram (see also Refs.\cite%
{AltshulerJETP1983}, \cite{Dorin-etal-PRB1993}):

\begin{equation*}
\sigma _{xx}^{LV\left( 7\right) }\simeq \frac{e^{2}}{4\hbar d}\frac{14\zeta
\left( 3\right) }{\pi ^{4}}\int \frac{d\left( \eta _{\left( 2\right)
}q^{2}\right) }{\varepsilon +\left( \eta _{\left( 2\right) }q^{2}\right) }=-%
\frac{1}{3}\sigma _{xx}^{LV\left( 5\right) }
\end{equation*}%
so that together with the identity: $\sigma _{xx}^{LV\left( 7\right)
}=\sigma _{xx}^{LV\left( 8\right) }$ one finds: 
\begin{eqnarray}
&\sigma _{xx}^{LV\left( 5+6+7+8\right) }=2\sigma _{xx}^{LV\left( 5+7\right)
}=  \notag \\
&-\frac{e^{2}}{\hbar d}\left( \frac{14\zeta \left( 3\right) }{\pi ^{4}}%
\right) \int \frac{d\left( \eta _{\left( 2\right) }q^{2}\right) }{%
\varepsilon +\left( \eta _{\left( 2\right) }q^{2}\right) }
\label{sig^LV(5-8)}
\end{eqnarray}

\subsection{Comparison with the TDGL functional approach}

\subsubsection{The Aslamazov-Larkin paraconductivity}

The TDGL functional $\mathfrak{L}\left( \Delta ,\mathbf{A}\right) $ of the
order parameter $\Delta \left( \boldsymbol{r},t\right) $ and vector
potential $\mathbf{A}\left( \boldsymbol{r},t\right) $ determines the
Cooper-pairs current density \cite{FuldeMaki70}: 
\begin{equation}
\mathbf{j}\left( \mathbf{r},t\right) =-c\frac{\partial \mathfrak{L}\left(
\Delta \left( \mathbf{r},t\right) ,\mathbf{A}\left( \mathbf{r},t\right)
\right) }{\partial \mathbf{A}\left( \mathbf{r},t\right) }  \label{j(r,t)}
\end{equation}%
responsible for the AL paraconductivity (see Appendix B).

In this approach the entire underlying information about the thin film of
pairing electrons system is incorporated in the inverse fluctuation
propagator (in wavevector-frequency representation) $D^{-1}\left( \mathbf{q+}%
2e\mathbf{A/\hbar },-i\Omega \right) $, mediating between the order
parameter and the GL functional. In the Gaussian approximation the relation
is quadratic, i.e.: 
\begin{eqnarray}
\mathfrak{L}\left( \Delta ,\mathbf{A}\right) &=&\frac{1}{d}\int \frac{d^{2}q%
}{\left( 2\pi \right) ^{2}}\left( \frac{1}{2\pi }\right) \int d\Omega \left
\vert \Delta \left( \mathbf{q},\Omega \right) \right \vert ^{2}\times  \notag
\\
&&\mathcal{D}^{-1}\left( \mathbf{q+}2e\mathbf{A/\hbar },-i\Omega \right)
\label{GLLag}
\end{eqnarray}%
so that the coupling to the external electromagnetic field takes place
directly through the vertex of the Cooper-pair current, defined in Eq.(\ref%
{j(r,t)}).

The corresponding AL time-ordered current-current correlator is given by: 
\begin{eqnarray}
&&Q_{AL}\left( i\Omega _{\nu }\right) =\left( 4eN_{2D}D\right)
^{2}d^{-1}\left( \frac{1}{2\pi }\right) ^{2}\int d^{2}qq_{x}^{2}k_{B}T
\label{Q_ALgen} \\
&&\sum \limits_{k=-\infty }^{\infty }C\left( q,\Omega _{k}+\Omega _{\nu
}\right) \mathcal{D}\left( q,\Omega _{k}+\Omega _{\nu }\right) C\left(
q,\Omega _{k}\right) \mathcal{D}\left( q,\Omega _{k}\right)  \notag
\end{eqnarray}%
where $\Omega _{k}=2kk_{B}T/\hbar ,\Omega _{\nu }=2\nu k_{B}T/\hbar ,$ $%
k=0,\pm 1,\pm 2,...,$\ \ \ \ $\nu =0,1,2,....$ are bosonic Matsubara
frequencies. Here, like in Sec.IIIA1, the electrical current is generated
along the $x$ axis, $q_{z},q_{x}$ are the fluctuation (in-plane) wave-vector
components along the magnetic and electric field directions, respectively,
and $q^{2}\equiv q_{z}^{2}+q_{x}^{2}$.

The fluctuation propagator $\mathcal{D}\left( q,\Omega _{k}\right) $ and its
corresponding effective current vertex $C\left( q,\Omega _{k}\right) $,
derived in Ref.\cite{MZPRB2021} without restriction to small values of $q$,
are given by:

\begin{eqnarray}
\mathcal{D}\left( q,\Omega _{k}\right) &=&\frac{1}{N_{2D}\Phi \left( x+\left
\vert k\right \vert /2\right) },  \notag \\
C\left( q,\Omega _{k}\right) &=&\frac{\Phi ^{\prime }\left( x+\left \vert
k\right \vert /2\right) }{4\pi k_{B}T}  \label{D&C}
\end{eqnarray}%
where for zero field: 
\begin{equation}
\Phi \left( x+\left \vert k\right \vert /2\right) =\varepsilon +\psi \left(
1/2+x+\left \vert k\right \vert /2\right) -\psi \left( 1/2\right)
\label{Phi}
\end{equation}%
and: $x=\hbar Dq^{2}/4\pi k_{B}T$. \ Here we note that $\mathcal{D}\left(
q,\Omega _{k}\right) $ in Eq.\ref{D&C} reduces to Eq.\ref{D(q,Omk)}
following the linearization of $\Phi \left( x+\left \vert k\right \vert
/2\right) $.

We now define a normalized effective current vertex analogous to $%
B_{x}\left( q,0,0\right) $: 
\begin{eqnarray}
C_{x}\left( q,\Omega _{k}\right) &\equiv &2N_{2D}Dq_{x}C\left( q,\Omega
_{k}\right) =2N_{2D}D\frac{\Phi ^{\prime }\left( x+\left \vert k\right \vert
/2\right) }{4\pi k_{B}T}q_{x}  \notag \\
&=&2N_{2D}D\frac{\psi ^{\prime }\left( \frac{1}{2}+\frac{\hbar \left \vert
\Omega _{k}\right \vert }{4\pi k_{B}T}+\frac{\hbar Dq^{2}}{4\pi k_{B}T}%
\right) }{4\pi k_{B}T}q_{x}  \label{C_xdef}
\end{eqnarray}%
with the help of which Eq.\ref{Q_ALgen} is rewritten in a form similar to Eq.%
\ref{Q^LV(1)}, i.e.: 
\begin{eqnarray}
&&Q_{AL}\left( i\Omega _{\nu }\right) \simeq 4e^{2}k_{B}T\sum
\limits_{k=-\infty }^{\infty }\frac{1}{d}\left( \frac{1}{2\pi }\right)
^{2}\times  \label{Q_AL} \\
&&\int d^{2}q\mathcal{D}\left( q,\Omega _{k}+\Omega _{\nu }\right)
C_{x}\left( q,\Omega _{k}\right) \mathcal{D}\left( q,\Omega _{k}\right)
C_{x}\left( q,\Omega _{k}\right)  \notag
\end{eqnarray}

Taking the static limit, $\Omega _{k}\rightarrow 0$ Eq.\ref{C_xdef} reduces
to:

\begin{eqnarray}
C_{x}\left( q,\Omega _{k}\rightarrow 0\right) &\rightarrow& 2N_{2D}D\frac{%
\psi ^{\prime }\left( \frac{1}{2}+\frac{\hbar Dq^{2}}{4\pi k_{B}T}\right) }{%
4\pi k_{B}T}q_{x}  \notag \\
&=& -B_{x}\left( q,0,0\right)  \label{C_x(q)}
\end{eqnarray}

The complete agreement between the expressions Eqs.\ref{B_x(q,0,0)} and \ref%
{C_x(q)}, clearly indicates that the effect of the infinite set of ladder
diagrams which renormalize the pairing vertices outside the fluctuation
propagators (Cooperon insertions) in the AL diagram, within the LV
microscopic approach, is fully consistent with the impurity-scatterings
effect introduced to the fluctuation current vertex through its relation to
the fluctuation propagator within our TDGL functional approach. The
corresponding consistency equation between the renormalized current vertex
and the fluctuation propagator, inherent to the TDGL functional approach,
should also be satisfied within the fully microscopic approach.

Within our TDGL functional approach we note that: $C_{x}\left( q,\Omega
_{k}\rightarrow 0\right) =\left( v_{F}^{2}\tau _{SO}/4\pi k_{B}T\right)
N_{2D}\Phi ^{\prime }\left( x\right) q_{x}$, whereas: $N_{2D}\Phi ^{\prime
}\left( x\right) =\left( 8\pi k_{B}T/\hbar v_{F}^{2}\tau _{SO}\right) \left[
\partial /\partial \left( q^{2}\right) \right] \mathcal{D}^{-1}\left(
q,0\right) $, so that:

\begin{equation}
C_{x}\left( q,0\right) =\frac{2q_{x}}{\hbar }\frac{\partial }{\partial
\left( q^{2}\right) }\mathcal{D}^{-1}\left( q,0\right) =-B_{x}\left(
q,0,0\right)  \label{ConsEqC}
\end{equation}

Equation \ref{ConsEqC} relates the (Cooperon) external-vertex insertions to
the internal-vertex insertions of ladder diagrams introduced in the
calculation of the fluctuation propagator. In our TDGL approach it appears
naturally, directly from the inverse fluctuation propagator, without taking
any additional measure. It reflects the fundamental variation principle,
satisfied by the electromagnetically modified GL free energy functional with
respect to the vector potential, see Eq.\ref{varGL_F} in Appendix B.

\subsubsection{The DOS conductivity}

The basic observable used in evaluating the DOS conductivity within the TDGL
approach is the CPFs density:

\begin{equation}
n_{CPF}=\frac{1}{d}\frac{1}{\left( 2\pi \right) ^{2}}\int \left \langle
\left \vert \phi \left( \mathbf{q}\right) \right \vert ^{2}\right \rangle
d^{2}q  \label{n_s}
\end{equation}%
with the Cooper-pair momentum distribution function $\left \langle
\left
\vert \phi \left( \mathbf{q}\right) \right \vert ^{2}\right \rangle $
derived by using the frequency-dependent GL functional, Eq.(\ref{GLLag}).
This is done by rewriting Eq.(\ref{GLLag}) in terms of the frequency and
wavenumber representations GL wavefunctions $\phi \left( \mathbf{q},\Omega
\right) $, after analytic continuation to real frequencies $i\Omega _{\mu
}\rightarrow \Omega $, i.e.:

\begin{eqnarray}
\mathfrak{L}\left( \Delta \right) &=&\frac{1}{d}\int \frac{d^{2}q}{\left(
2\pi \right) ^{2}}\int \frac{d\Omega }{2\pi }\left \vert \Delta \left( 
\mathbf{q},\Omega \right) \right \vert ^{2}\mathcal{D}^{-1}\left( q,-i\Omega
\right)  \label{Lag(phi)} \\
&=&\frac{1}{d}\int \frac{d^{2}q}{\left( 2\pi \right) ^{2}}\int \frac{d\Omega 
}{2\pi }\left \vert \phi \left( \mathbf{q},\Omega \right) \right \vert
^{2}L^{-1}\left( q,\Omega \right) =\mathfrak{L}\left( \phi \right)  \notag
\end{eqnarray}%
under the normalization relations:

\begin{equation}
\mathcal{D}\left( q,-i\Omega \right) =\left( \frac{\alpha k_{B}T}{N_{2D}}%
\right) L\left( q,\Omega \right)  \label{NormalizRelat}
\end{equation}

Here the constant $\alpha $ (i.e. independent of $q$ and $\Omega $),
determines the normalization of the wavefunctions from the corresponding
components of the order parameter through the reciprocal relations (see
Appendix C):

\begin{equation*}
\left \vert \phi \left( \mathbf{q},\Omega \right) \right \vert ^{2}=\left( 
\frac{N_{2D}}{\alpha k_{B}T}\right) \left \vert \Delta \left( \mathbf{q}%
,\Omega \right) \right \vert ^{2}
\end{equation*}

Following LV, the resulting expression for the momentum distribution
function is obtained by exploiting the Langevin force technique in the TDGL
equation, which leads to:

\begin{equation*}
\left \langle \left \vert \phi \left( \mathbf{q}\right) \right \vert
^{2}\right \rangle =2k_{B}T\gamma _{GL}\int \frac{d\left( \hbar \Omega
\right) }{2\pi }\left \vert L\left( q,\Omega \right) \right \vert ^{2}
\end{equation*}%
where for small $q$ values:

\begin{equation}
L\left( q,\Omega \right) ^{-1}=k_{B}T\alpha \left( \varepsilon +\xi \left(
T\right) ^{2}q^{2}\right) -i\gamma _{GL}\hbar \Omega  \label{L(q)^-1}
\end{equation}%
with the CPF coherence length $\xi \left( T\right) $ and dimensionless
life-time parameter, $\gamma _{GL}$, are given by: 
\begin{equation}
\xi \left( T\right) =\sqrt{\frac{\pi \hbar }{8k_{B}T}D},\gamma _{GL}=\frac{%
\pi \alpha }{8}  \label{xi,gam_GL}
\end{equation}

Note that the actual CPF life-time can be found from the pole of the GL
propagator $L\left( q,\Omega \right) $ to be: 
\begin{equation}
\tau _{GL}\left( q\right) =\frac{\gamma _{GL}}{k_{B}T\alpha \left(
\varepsilon +\xi \left( T\right) ^{2}q^{2}\right) }  \label{tau_GL(q)}
\end{equation}

Performing the frequency integration we find for the momentum distribution
function:

\begin{equation}
\left \langle \left \vert \phi \left( \mathbf{q}\right) \right \vert
^{2}\right \rangle =\frac{1}{\alpha }\frac{1}{\varepsilon +\xi \left(
T\right) ^{2}q^{2}}  \label{momentdistrH0}
\end{equation}

At this point one notes that Eq.\ref{sig^LV(5-8)} for the LV DOS
conductivity has the basic structure of an effective Drude formula,
originally proposed by LV, and more recently used within the TDGL approach
in Ref.\cite{MZJPC2023}, that is: 
\begin{eqnarray}
&&\sigma _{xx}^{LV\left( 5+6+7+8\right) }\propto -\left( \frac{e^{2}}{d\hbar 
}\right) \left( \frac{4}{\pi ^{2}}\right) \frac{k_{B}T}{E_{F}}\frac{1}{%
\alpha }\int \frac{d\left( \xi \left( T\right) ^{2}q^{2}\right) }{%
\varepsilon +\xi \left( T\right) ^{2}q^{2}}  \notag \\
&=&-2n_{CPF}\frac{e^{2}}{m^{\ast }}\tau _{SO}\equiv \sigma _{DOS}^{TDGL}
\label{sig_DOSH0}
\end{eqnarray}%
where $n_{s}$ is given by Eq.\ref{n_s} and the momentum distribution
function by Eq.\ref{momentdistrH0}. Note also that Eq.\ref{sig^LV(5-8)}
originates in two equal contributions from two groups of two diagrams shown
in Fig.1; diagrams $\left( 5,7\right) $, and diagrams $\left( 6,8\right) $.
Thus, identifying Eq.\ref{sig_DOSH0} with the basic contribution to the LV
DOS conductivity, that is:

\begin{equation}
\sigma _{DOS}^{TDGL}=\sigma _{xx}^{LV\left( 5+7\right) }=\sigma
_{xx}^{LV\left( 6+8\right) }  \label{sig_xx^(5+7)}
\end{equation}%
the normalization constant should read:

\begin{equation}
\alpha =\frac{4\pi ^{2}}{7\zeta \left( 3\right) }\frac{k_{B}T}{E_{F}}
\label{alpha}
\end{equation}

This is exactly the expression for $\alpha $ obtained in the clean limit
(i.e. for $\xi \left( T\right) =\sqrt{7\zeta \left( 3\right) /8}\left( \hbar
v_{F}/2\pi k_{B}T\right) $) by requiring the GL propagator in Eq.\ref%
{L(q)^-1} to have the Schrodinger-like form with the Cooper-pair mass equals
twice the free electron mass (see Appendix C). In the dirty-limit situation
under study here one therefore evaluates the momentum distribution function
in Eq.\ref{momentdistrH0} with the dirty limit coherence length, $\xi \left(
T\right) =\sqrt{\pi \hbar D/8k_{B}T}$, and $\alpha $ given by Eq.\ref{alpha}%
, since normalization of the wavefunctions should be independent of the
effect of scatterings.

It should also be noted that $\sigma _{xx}^{LV\left( 5+7\right) }$ has been
derived within the microscopic LV approach while neglecting the $q$
dependence of the renormalized pairing vertex factor (see Eq.\ref{lambda(q)}%
). Inclusion of this dependence would transform Eq. \ref{sig^LV(5-8)} to:

\begin{eqnarray}
&\sigma _{DOS}^{LV} =2\sigma _{xx}^{LV\left( 5+7\right) }
\label{sig_DOS^LVq} \\
&=\left( \frac{e^{2}}{d\hbar }\right) \left( \frac{1}{\pi ^{4}}\right) \int
d\left( \eta _{\left( 2\right) }q^{2}\right) \frac{\psi ^{\prime \prime
}\left( \frac{1}{2}+\frac{\hbar D}{4\pi k_{B}T}q^{2}\right) }{\varepsilon
+\eta _{\left( 2\right) }q^{2}}  \notag
\end{eqnarray}

In this expression, due to the fact that $\alpha $ is independent of $q$,
the CPFs momentum distribution function Eq.\ref{momentdistrH0}, derived
above by neglecting the $q$ dependence of the (Cooperon) factor $\psi
^{\prime \prime }\left( 1/2+\hbar Dq^{2}/4\pi k_{B}T\right) $, is clearly
identified under the integral, separable from any additional $q$ dependent
factors, so that $\sigma _{xx}^{LV\left( 5+7\right) }$can be written in the
momentum dependent Drude-like form:

\begin{equation}
\sigma _{xx}^{LV\left( 5+7\right) }=-2\frac{e^{2}}{m^{\ast }}\frac{1}{d}\int 
\frac{d^{2}q}{\left( 2\pi \right) ^{2}}\left \langle \left \vert \phi \left(
q\right) \right \vert ^{2}\right \rangle \tau _{SO}\left( q\right)
\label{sig_DOS^LVmod}
\end{equation}

Here the additional $q$ dependence appears as an effective correction to the
single-electron relaxation time: 
\begin{eqnarray}
&&\tau _{SO}\left( q\right) \equiv \tau _{SO}\frac{\psi ^{\prime \prime
}\left( \frac{1}{2}+\frac{\hbar D}{4\pi k_{B}T}q^{2}\right) }{\psi ^{\prime
\prime }\left( \frac{1}{2}\right) },  \label{tau_SO(q)} \\
&&\tau _{SO}\left( q\rightarrow 0\right) =\tau _{SO}  \notag
\end{eqnarray}

The resulting expression for $\tau _{SO}\left( q\right) $ shows that the
relaxation time of electrons involved in scattering by CPFs (see Fig.2) is
effectively suppressed at any exchange of momentum ( $q>0$ ) with a CPF.

Note that Eq.\ref{sig_DOS^LVmod}, without the $q$ dependence of $\tau
_{SO}\left( q\right) $ is well defined in the zero temperature limit. The
result, after integration over $q$ with the cutoff $q_{c}=1/\xi \left(
T\right) $ (see Sec.V for discussion of the cutoff), has a nonvanishing
value, that is:\ 

\begin{equation}
\sigma _{xx}^{LV\left( 5+7\right) }\simeq -\left( \frac{7\zeta \left(
3\right) }{\pi ^{4}}\right) \left( \frac{e^{2}}{d\hbar }\right) \ln \left( 1+%
\frac{1}{\varepsilon }\right)  \label{sig_DOS^LVq0}
\end{equation}

Taking into account the $q$ dependence of $\tau _{SO}\left( q\right) $ (Eq.%
\ref{tau_SO(q)}) under the integration over $q$ in Eq.\ref{sig_DOS^LVmod}
significantly suppresses $\sigma _{DOS}^{LV}$ at very low temperature,
however, its non vanishing zero temperature limit (see Appendix D) is due to
quantum fluctuations.

\section{ Extension of the microscopic theory to finite magnetic field}

Considering, either Eq.\ref{Q^LV(1)} for the current-current correlator of
the AL diagram, or Eq.\ref{Q^LV(5)} for the correlator of the 5-th LV
diagram, the new ingredient associated with the finite magnetic field at low
temperatures, besides the important modification introduced to the
fluctuation propagator (see below), is the renormalized pairing vertex
factor, which includes the effect of spin-orbit scatterings and the magnetic
field effect through Zeeman spin splitting $\mu _{B}H=\hslash I$ (see Ref.%
\cite{MZPRB2021}):

\begin{equation}
\lambda _{\pm }\left( q,\omega _{n},-\omega _{n}\right) \equiv \frac{S_{\pm
}\left( \omega _{n},q\right) }{S_{\pm }^{0}\left( \omega _{n},q\right) }
\label{lambda_pm-stand}
\end{equation}%
where:%
\begin{equation*}
S_{\pm }^{0}\left( \omega _{n}\mathbf{,}q\right) \approx \frac{\pi }{\left
\vert \omega _{n}\right \vert +\frac{1}{2\tau _{SO}}+\frac{1}{2}Dq^{2}\pm
iIsign\left( \omega _{n}\right) },
\end{equation*}%
and:

\begin{equation*}
S_{\pm }\left( \omega _{n}\mathbf{,}q\right) \approx \pi \frac{\left \vert
\omega _{n}\right \vert +\frac{1}{\tau _{SO}}+\frac{1}{2}Dq^{2}\mp
iIsign\left( \omega _{n}\right) }{\left( \left \vert \omega _{n}\right \vert
+\frac{1}{2\tau _{SO}}+\frac{1}{2}Dq^{2}\right) ^{2}-\left( \frac{1}{2\tau
_{SO}}\right) ^{2}+I^{2}}
\end{equation*}%
so that: 
\begin{widetext}
\begin{equation}
\lambda _{\pm }\left( q,\omega _{n},-\omega _{n}\right) \approx \frac{\left[
\left \vert \omega _{n}\right \vert +\frac{1}{\tau _{SO}}+\frac{1}{2}%
Dq^{2}\mp iIsgn\left( \omega _{n}\right) \right] \left[ \left \vert \omega
_{n}\right \vert +\frac{1}{2\tau _{SO}}\pm iIsgn\left( \omega _{n}\right) +%
\frac{1}{2}Dq^{2}\right] }{\left( \left \vert \omega _{n}\right \vert +\frac{%
1}{2}Dq^{2}\right) \left( \left \vert \omega _{n}\right \vert +\frac{1}{\tau
_{SO}}+\frac{1}{2}Dq^{2}\right) +I^{2}}  \label{lam_qH}
\end{equation}
\end{widetext}

For the zero-field case, $I=0$, we recover LV result, Eq.\ref{lambda(q)}.

It is seen to correspond to coherent scattering on the same impurity by a
pair of free electrons, entering to or emerging from states of fluctuating
Cooper-pairs \cite{Lopatinetal05},\cite{GalitLarkinPRB01}.

Another important modification due to the presence of the magnetic field is
introduced to the fluctuation propagator. An explicit expression, e.g. for
the static propagator derived in Ref.\cite{MZPRB2021}, takes the form:

\begin{equation}
\mathcal{D}^{H}\left( q\right) =\frac{1}{N_{2D}\Phi \left( q;H\right) }
\label{D^H(q)}
\end{equation}%
where:

\begin{eqnarray}
&&\Phi \left( q;H\right) =\varepsilon _{H}+a_{+}\left[ \psi \left( \frac{1}{2%
}+f_{-}+x\right) -\psi \left( \frac{1}{2}+f_{-}\right) \right]  \notag \\
&&+a_{-}\left[ \psi \left( \frac{1}{2}+f_{+}+x\right) -\psi \left( \frac{1}{2%
}+f_{+}\right) \right]  \label{Phi(x;H)}
\end{eqnarray}%
and $x=\hbar Dq^{2}/4\pi k_{B}T$. \ In this expression: 
\begin{equation}
\varepsilon _{H}\equiv \ln \left( \frac{T}{T_{c0}}\right) +a_{+}\psi \left( 
\frac{1}{2}+f_{-}\right) +a_{-}\psi \left( \frac{1}{2}+f_{+}\right) -\psi
\left( \frac{1}{2}\right)  \label{eps_H}
\end{equation}%
is the critical shift parameter, and: 
\begin{eqnarray*}
a_{\pm } &=&\frac{1}{2}\left \{ 1\pm \left[ 1-\left( \frac{2\mu _{B}H}{%
\varepsilon _{SO}}\right) ^{2}\right] ^{-1/2}\right \} , \\
f_{\pm } &=&\frac{\varepsilon _{SO}}{4\pi k_{B}T}\left \{ 1\pm \left[
1-\left( \frac{2\mu _{B}H}{\varepsilon _{SO}}\right) ^{2}\right]
^{1/2}\right \}
\end{eqnarray*}

Eq.\ref{D^H(q)}, with Eq.\ref{Phi(x;H)}, for the static fluctuation
propagator at finite field, is usually approximated (at small $q$ values),
by the linear expansion with respect to the kinetic energy, that is:

\begin{equation}
\mathcal{D}^{H}\left( q\right) \simeq \frac{1}{N_{2D}}\frac{1}{\varepsilon
_{H}+\eta \left( H\right) \frac{\hbar Dq^{2}}{4\pi k_{B}T}}  \label{D(q)^H}
\end{equation}%
where:\ 
\begin{equation}
\eta \left( H\right) =a_{+}\psi ^{\prime }\left( \frac{1}{2}+f_{-}\right)
+a_{-}\psi ^{\prime }\left( \frac{1}{2}+f_{+}\right)  \label{eta(H)}
\end{equation}%
is the field-dependent reduced stiffness coefficient of the fluctuations
modes. At zero field, where $\eta \left( H=0\right) =\psi ^{\prime }\left( 
\frac{1}{2}\right) =\pi ^{2}/2$, Eq.\ref{D(q)^H} reduces to the static limit
of Eq.\ref{D(q,Omk)}. At finite field, $\eta \left( H\right) $ tends to zero
with vanishing temperature which effectively leads to complete uniform
softening of all fluctuation modes. This remarkable field-induced softening
takes place under suppressed pair-breaking condition, $2\mu _{B}H<$ $\hbar
/\tau _{SO}$, i.e. within a fields range restricted by the spin-orbit
scattering rate. The strong spin-orbit scattering relevant to our paper,
i.e. typically with $\tau _{SO}\approx 2\times 10^{-13}\boldsymbol{\mathit{s}%
}$, is of an order of magnitude rate larger than $\mu _{B}H_{c\parallel
0}/\hbar $. Thus, the softening effect could be realized even in the more
usual situations of relatively weak spin-orbit scatterings.

\subsection{The Aslamazov-Larkin paraconductivity diagram}

Starting with Eq.\ref{Q^LV(1)} for the current-current correlator of the AL
diagram, following the approximation in which the boson frequency arguments
of the effective current vertex $B_{x}\left( q,\Omega _{k},\Omega _{\nu
}\right) $ are set to zero, the latter for a finite magnetic field $H$, can
be written in the simplified form:

\begin{eqnarray}
&&B_{x}^{H}\left( q,\Omega _{k}=0,\Omega _{\nu }=0\right) \simeq -\frac{1}{2}%
\pi N_{2D}\frac{1}{2}v_{F}^{2}q_{x}\times  \notag \\
&&k_{B}T\sum_{n}\frac{\lambda _{+}^{H}\left( 0,\omega _{n},-\omega
_{n}\right) \lambda _{-}^{H}\left( 0,\omega _{n},-\omega _{n}\right) }{\left[
\left \vert \widetilde{\omega }_{n}\right \vert -iIsign\left( \omega
_{n}\right) \right] ^{3}}  \label{B_xH}
\end{eqnarray}%
and in which we use the $q\rightarrow 0$ \ limit of the renormalized pairing
vertex, Eq.\ref{lam_qH}.

Using the above expression for the effective current vertex, we may write
the corresponding correlator in the form:%
\begin{eqnarray*}
&&Q_{xx}^{AL}\left( i\Omega _{\nu }\right) =-4e^{2}k_{B}T\frac{1}{d}\int 
\frac{d^{2}q}{\left( 2\pi \right) ^{2}} \\
&&\times \left[ B_{x}^{H}\left( q,\Omega _{k}=0,\Omega _{\nu }=0\right) %
\right] ^{2}\Psi \left( q\mathbf{,}\Omega _{\nu }\right)
\end{eqnarray*}%
where: 
\begin{equation*}
\Psi \left( q\mathbf{,}\Omega _{\nu }\right) =\sum \limits_{k}\mathcal{D}%
\left( q,\Omega _{\nu }+\Omega _{k}\right) \mathcal{D}\left( q,\Omega
_{k}\right)
\end{equation*}

Carrying out the boson frequency summation and the analytic continuation: $%
i\Omega _{\nu }\rightarrow \omega $, we obtain after expansion in small $%
\omega $, to first order:

\begin{equation*}
\Psi \left( q\mathbf{,}-i\omega \right) -\Psi \left( q\mathbf{,}0\right)
\rightarrow \frac{1}{N_{2D}^{2}}\frac{\frac{i\hbar \omega }{4\pi k_{B}T}\psi
^{\prime }\left( \frac{1}{2}+\frac{\left( \mu _{B}H\right) ^{2}}{2\pi
k_{B}T\varepsilon _{SO}}\right) }{\left[ \varepsilon _{H}+\eta \left(
H\right) \frac{\hbar Dq^{2}}{4\pi k_{B}T}\right] ^{3}}
\end{equation*}%
in which the linear approximation, Eq.\ref{D(q)^H} for the static
fluctuation propagator is used.

Complementing the expression for $\lambda _{\pm }\left( q=0,\omega
_{n},-\omega _{n}\right) $ in Eq.\ref{lam_qH} with the dirty limit condition 
$\left( \tau _{SO}\mu _{B}H/\hbar \right) ^{2}\ll 1$, we find that: 
\begin{equation}
\lambda _{+}^{H}\left( 0,\omega _{n},-\omega _{n}\right) \lambda
_{-}^{H}\left( 0,\omega _{n},-\omega _{n}\right) \approx \frac{\widetilde{%
\omega }_{n}^{2}}{\left( \left \vert \omega _{n}\right \vert +\tau
_{SO}I^{2}\right) ^{2}}  \label{lamPlamM}
\end{equation}%
and the corresponding sheet paraconductivity: 
\begin{eqnarray*}
&&\sigma _{xx}^{AL}\left( \omega \rightarrow 0\right) =\lim_{\omega
\rightarrow 0}\frac{i}{\omega }\left[ Q_{xx}^{AL}\left( \omega \right)
-Q_{xx}^{AL}\left( 0\right) \right] \\
&=&-4e^{2}k_{B}T\frac{1}{d}\int \frac{d^{2}q}{\left( 2\pi \right) ^{2}}\left[
B_{x}^{H}\left( q,\Omega _{k}=0,\Omega _{\nu }=0\right) \right] ^{2} \\
&&\times \lim_{\omega \rightarrow 0}\frac{i}{\omega }\left[ \Psi \left( q%
\mathbf{,}-i\omega \right) -\Psi \left( q\mathbf{,}0\right) \right]
\end{eqnarray*}%
, is written in the form: 
\begin{eqnarray}
\sigma _{xx}^{AL}\left( \omega \rightarrow 0\right) &\simeq &e^{2}\frac{1}{%
2\pi d}\int \frac{d^{2}q}{\left( 2\pi \right) ^{2}}\left( B_{x}^{H}\left(
q,\Omega _{k}=0,\Omega _{\nu }=0\right) \right) ^{2}  \notag \\
&\times &\frac{1}{N_{2D}^{2}}\frac{\hbar \eta _{LV}\left( H\right) }{\left(
\varepsilon _{H}+\eta \left( H\right) \frac{\hbar Dq^{2}}{4\pi k_{B}T}%
\right) ^{3}}  \label{sig_xx^AL}
\end{eqnarray}%
where the effective current vertex is given by: 
\begin{equation}
B_{x}^{H}\left( q,\Omega _{k}=0,\Omega _{\nu }=0\right) =N_{2D}\frac{D}{2\pi
k_{B}T}\eta _{LV}\left( H\right) q_{x}  \label{B_x^H}
\end{equation}%
with the (dirty limit) LV version of the reduced stiffness function: 
\begin{equation}
\eta _{LV}\left( H\right) \simeq \psi ^{\prime }\left( \frac{1}{2}+\frac{1}{2%
}\frac{\left( \mu _{B}H\right) ^{2}}{\pi k_{B}T\varepsilon _{SO}}\right)
\label{eta_LV}
\end{equation}

Using Eq.\ref{B_x^H}, Eq.\ref{sig_xx^AL} reduces to: {\small 
\begin{equation*}
\sigma _{AL}^{mic}\left( H\right) \approx e^{2}\frac{1}{d\hbar }\frac{1}{%
2\pi ^{2}}\eta _{LV}\left( H\right) \left[ \frac{\eta _{LV}\left( H\right) }{%
\eta \left( H\right) }\right] ^{2}\int_{0}^{\widetilde{x}_{c}}d\widetilde{x}%
\frac{\widetilde{x}}{\left( \varepsilon _{H}+\widetilde{x}\right) ^{3}}
\end{equation*}%
} where $\widetilde{x}_{c}=\eta \left( H\right) x_{c},x_{c}=\hbar
Dq_{c}^{2}/4\pi k_{B}T$, and $q_{c}$ is the cutoff wavenumber.

It is easy to check (see also Appendix E) that under the dirty limit
conditions: $4\pi k_{B}T\tau _{SO}/\hbar ,2\mu _{B}H\tau _{SO}/\hbar \ll 1$: 
\begin{equation}
\eta \left( H\right) \approx \eta _{LV}\left( H\right) \approx \psi ^{\prime
}\left( \frac{1}{2}+\frac{1}{2}\frac{\left( \mu _{B}H\right) ^{2}}{\pi
k_{B}T\varepsilon _{SO}}\right)  \label{eta(H)approx}
\end{equation}%
so that finally we find for the AL diagram conductivity at finite field::

\begin{equation}
\sigma _{AL}^{mic}\left( H\right) \approx \frac{1}{16}\left( \frac{e^{2}}{%
d\hbar }\right) \frac{\widetilde{\eta }\left( H\right) }{\varepsilon
_{H}\left( 1+\frac{\varepsilon _{H}}{\eta \left( H\right) x_{c}}\right) }
\label{sig_AL(H)^mic}
\end{equation}%
where: 
\begin{equation}
\widetilde{\eta }\left( H\right) \equiv \frac{\eta \left( H\right) }{\eta
\left( 0\right) }=\frac{2}{\pi ^{2}}\eta \left( H\right)
\label{tilde(eta(H))}
\end{equation}

This result is in full agreement with the AL conductivity obtained within
the TDGL functional approach (see in Appendix F an erratum of the original
derivation presented in Ref.\cite{MZJPC2023}). This result is expected, of
course, in light of the equivalence between the TDGL functional approach and
the microscopic LV (diagrammatic) theory applied to the AL paraconductivity
calculation, as discussed in detailed in Sec.III.

\bigskip

\subsection{The DOS conductivity diagrams}

Starting with Eq.\ref{Q^LV(5)} for the current-current correlator of the
5-th LV diagram, we use, following LV, the approximate expression for the
electronic kernel: 
\begin{widetext}
\begin{equation*}
\Sigma _{xx}^{\left( 5\right) H}\left( q\rightarrow 0,\Omega _{k}\rightarrow
0,\Omega _{\nu }\right) =k_{B}T\sum \limits_{n}\lambda _{+}^{H}\left(
0,\omega _{n},-\omega _{n}\right) \lambda _{-}^{H}\left( 0,\omega
_{n},-\omega _{n}\right) I_{xx}^{\left( 5\right) }\left( 0,\omega
_{n},0,\Omega _{\nu }\right)
\end{equation*}%
with the product of the dirty-limit renormalized pairing vertex functions,
Eq.\ref{lamPlamM}, arriving at the expression:

\begin{eqnarray}
&\Sigma _{xx}^{\left( 5\right) H}\left( q\rightarrow 0,\Omega
_{k}\rightarrow 0,\Omega _{\nu }\right)  \label{Sig_xx(5)H} 
\rightarrow -\pi N_{2D}v_{F}^{2}\frac{k_{B}T}{\hbar ^{3}}\times  \notag \\
&\left[ 
\left( \sum \limits_{n=-\infty }^{-\nu -1}+\sum \limits_{n=0}^{\infty
}\right) \frac{1}{\left( 2\left \vert \omega _{n}\right \vert +2\tau
_{SO}I^{2}\right) ^{2}}\frac{1}{\widetilde{\omega }_{n}+\widetilde{\omega }
_{n+\nu }}sign\left( \omega _{n}\right) 
+\sum \limits_{n=-\nu }^{-1}\frac{\left( 2\widetilde{\omega }_{n}\right)
^{2} }{\left( 2\left \vert \omega _{n}\right \vert +2\tau _{SO}I^{2}\right)
^{2}} \frac{1}{\widetilde{\omega }_{n}+\widetilde{\omega }_{n+\nu }}\left( 
\frac{1 }{\left( \omega _{\nu }+1/\tau \right) ^{2}}-\frac{1}{\left( 2%
\widetilde{ \omega }_{n}\right) ^{2}}\right)%
\right]  
\end{eqnarray}

This expression is identical to the corresponding result for $\Sigma
_{xx}^{\left( 5\right) }$ in Ref.\cite{LV05} if one replaces in Eq.\ref%
{Sig_xx(5)H} the term $2\tau _{SO}I^{2}$\textbf{\ }with\textbf{\ }$2\times
\left( \frac{1}{2}Dq^{2}\right) $. The former term corresponds to the Zeeman
energy transfer, $2I=2\mu _{B}H/\hbar $, in the two-electron scattering
process occurring at each pairing vertex, whereas the latter corresponds to
the kinetic energy transfer close to the Fermi surface, $\mathbf{v}_{F}\cdot 
\mathbf{q}$ , taking place in such a scattering process.

The combined effect of the Zeeman and kinetic energy transfers on the
renormalized pairing vertex factor can be inferred from Eq.\ref{lam_qH} by
rewriting it in an approximate form:

\begin{equation}
\lambda _{\pm }^{H}\left( q,\omega _{n},-\omega _{n}\right) \approx \frac{1}{
\tau _{SO}}\frac{1}{2\left \vert \omega _{n}\right \vert +Dq^{2}+2\tau
_{SO}I^{2}}  \label{lamb^H(q,om)}
\end{equation}%
which has been derived under the dirty limit conditions: $1/\tau _{SO}\gg
2\tau _{SO}I^{2},Dq^{2},k_{B}T/\hbar .$ This expression reflects the dual
effect of the imbalance kinetic and magnetic energies in removing the zero
temperature singularity of the renormalized pairing vertex factor $\lambda
_{\pm }^{H}\left( q,\omega _{n},-\omega _{n}\right) $, which takes place
equivalently in the orbital and spin spaces.

Performing the fragmented Matsubara frequency summations and the analytic
continuation: $\Omega _{\nu }\rightarrow -i\omega $, we find:

\begin{eqnarray}
&&\Sigma _{xx}^{\left( 5\right) H}\left( 0,0,\Omega _{\nu }\rightarrow
-i\omega \right) -\Sigma _{xx}^{\left( 5\right) H}\left( 0,0,0\right)
\label{dSig_xx^(5)H} \\
&\approx &-i\omega \frac{N_{2D}v_{F}^{2}\tau _{SO}^{2}}{8\pi k_{B}T\hbar }%
\left[ \psi ^{\prime }\left( \frac{1}{2}+\frac{\hbar }{4\pi k_{B}T\tau _{SO}}%
\right) -\frac{3\hbar }{4\pi k_{B}T\tau _{SO}}\psi ^{\prime \prime }\left( 
\frac{1}{2}+\frac{\left( \mu _{B}H\right) ^{2}}{2\pi k_{B}T\varepsilon _{SO}}%
\right) \right]  \notag
\end{eqnarray}%
in which the first term within the square brackets can be neglected in the
dirty limit: $\hbar /4\pi k_{B}T\tau _{SO}\gg 1$.

Adding the complementary contribution of the 7-th diagram, and exploiting
the dirty-limit approximation mentioned above, we have for the combined
retarded current-current correlator:

\begin{equation}
Q_{xx}^{\left( 5+7\right) H}\left( i\Omega _{\nu }\rightarrow \omega \right)
\rightarrow -i\omega \frac{e^{2}}{d}\frac{D}{4\pi ^{2}k_{B}T}\psi ^{\prime
\prime }\left( \frac{1}{2}+\frac{\left( \mu _{B}H\right) ^{2}}{2\pi
k_{B}T\varepsilon _{SO}}\right)\int \frac{d^{2}q}{\left( 2\pi \right) ^{2}}%
\frac{1}{\varepsilon _{H}+\frac{\eta \left( H\right) \hbar D}{4\pi k_{B}T}%
q^{2}}  \label{Q_xx^(5+7)H}
\end{equation}

Adding the contributions of the 6-th plus 8-th diagrams, which are identical
to those of the 5-th and the 7-th ones respectively, we find:

\begin{equation}
\sigma _{xx}^{\left( 5+6+7+8\right) H}=2\sigma _{xx}^{\left( 5+7\right)
H}\simeq e^{2}\frac{1}{d\hbar }\frac{1}{\pi ^{4}}\psi ^{\prime \prime
}\left( \frac{1}{2}+\frac{\left( \mu _{B}H\right) ^{2}}{2\pi
k_{B}T\varepsilon _{SO}}\right) \int d\left( \eta _{\left( 2\right)
}q^{2}\right) \frac{1}{\varepsilon _{H}+\widetilde{\eta }\left( H\right)
\left( \eta _{\left( 2\right) }q^{2}\right) }  \label{sig_DOSH}
\end{equation}

Performing the integration the DOS conductivity at finite field is written
as:

\begin{equation}
\sigma _{DOS}^{mic}\left( H\right) \simeq -\left( \frac{e^{2}}{2\pi ^{2}}%
\right) \frac{1}{d\hbar }\frac{\left \vert \psi ^{\prime \prime }\left( 
\frac{1}{2}+\frac{\left( \mu _{B}H\right) ^{2}}{2\pi k_{B}T\varepsilon _{SO}}%
\right) \right \vert }{\eta \left( H\right) }\ln \left( 1+\frac{\eta \left(
H\right) x_{c}}{\varepsilon _{H}}\right)  \label{sig_DOSHfin}
\end{equation}%
with the cutoff: $x_{c}=\hbar Dq_{c}^{2}/4\pi k_{B}T$ 
\end{widetext},

\section{The Cooper-pair fluctuations density at finite field}

The detailed analysis presented in Sec.III indicates that Eq.\ref%
{momentdistrH0} for the CPFs momentum distribution function at zero field,
derived within the (exclusive boson) TDGL functional approach, can be
identified in the microscopic diagrammatic LV theory, while writing the
basic $(5+7)$-diagrams contribution to the DOS conductivity in a Drude-like
form, Eq.\ref{sig_DOS^LVmod}, with a $q$ dependent single-electron
relaxation time $\tau _{SO}\left( q\right) $ (see Eq.\ref{tau_SO(q)}). In
the presence of a finite magnetic field, while neglecting the $q$ dependence
of the renormalized pairing vertex factor (Eq.\ref{lamb^H(q,om)}), the
microscopic expression for the DOS conductivity, Eq.\ref{sig_DOSH}, includes
a field-dependent Cooperon factor, $\psi ^{\prime \prime }\left( 1/2+\left(
\mu _{B}H\right) ^{2}/2\pi k_{B}T\varepsilon _{SO}\right) $, which reflects
loss of single electron coherence by the magnetic field, analogous to the
corresponding zero-field $q$ dependent factor appearing in Eq.\ref%
{sig_DOS^LVq}. In the presence of both the $q$-dispersion and the magnetic
field, the loss of coherence originates at the pairing vertices shown in
Fig.2 for the DOS conductivity diagram, where pairs of single-electron lines
with nonzero center-of-mass momentum ($q\neq 0$) and total magnetic moment ($%
I\neq 0$) create and then annihilate CPFs. Physically speaking, it is
interpreted as suppression of electron coherence-time due to electron
scattering with background electrons via virtual exchange of CPFs.

\subsection{A proper definition of the Cooper-pair fluctuations density}

Repeating the calculation of Sec.IV.B by including the $q$ dependence of the
renormalized pairing vertex factor, Eq.\ref{lamb^H(q,om)}, the result for
the DOS conductivity in the presence of magnetic field: 
\begin{equation}
\sigma _{DOS}^{mic}\left( H\right) =2\sigma _{xx}^{LV\left( 5+7\right)
}\left( H\right)  \label{sig_DOS(H)mod}
\end{equation}%
can be written in a Drude-like form:

\begin{equation}
\sigma _{xx}^{LV\left( 5+7\right) }\left( H\right) =-2\frac{e^{2}}{m^{\ast }}%
\frac{1}{d}\int \frac{d^{2}q}{\left( 2\pi \right) ^{2}}\left \langle \left
\vert \phi \left( \mathbf{q}\right) \right \vert ^{2}\right \rangle _{H}\tau
_{SO}\left( q;H\right)  \label{sig_xx^(5+7)(H)}
\end{equation}%
which includes both the field and $q$ dependencies of the Cooperon factor,
i.e.: 
\begin{equation}
\tau _{SO}\left( q;H\right) \equiv \frac{\psi ^{\prime \prime }\left( \frac{1%
}{2}+\frac{\hbar D}{4\pi k_{B}T}q^{2}+\frac{\left( \mu _{B}H\right) ^{2}}{%
2\pi k_{B}T\varepsilon _{SO}}\right) }{\psi ^{\prime \prime }\left( \frac{1}{%
2}\right) }\tau _{SO}  \label{tau_SO(q;H)}
\end{equation}%
and the momentum distribution function: 
\begin{equation}
\left \langle \left \vert \phi \left( \mathbf{q}\right) \right \vert
^{2}\right \rangle _{H}\equiv \frac{1}{\alpha }\frac{1}{\varepsilon _{H}+%
\widetilde{\eta }\left( H\right) \left( \eta _{\left( 2\right) }q^{2}\right) 
}  \label{momdistrH}
\end{equation}%
with $\alpha $ given by Eq.\ref{alpha}.

Similar to the zero-field case discussed in Sec.III (see Ref.\cite{LV05}),
the TDGL-Langevin approach will be used below to derive Eq.\ref{momdistrH}
for the momentum distribution function in a magnetic field. The required
modification with respect to the zero-field case is associated with the
effect of the magnetic field on the fluctuating pairs life-time (given by Eq.%
\ref{tau_GL(q)} in the zero-field case). Thus, starting with the
TDGL-Langevin equation:

\begin{equation*}
\widehat{L}^{-1}\phi \left( \mathbf{r},t\right) =\zeta \left( \mathbf{r}%
,t\right)
\end{equation*}%
under the white-noise condition of the Langevin force correlator:%
\begin{equation}
\left \langle \zeta ^{\ast }\left( \mathbf{r},t\right) \zeta \left( \mathbf{r%
}^{\prime },0\right) \right \rangle =2k_{B}T\hbar \gamma _{GL}\left(
H\right) \delta \left( \mathbf{r}-\mathbf{r}^{\prime }\right) \delta \left(
t\right)  \label{LFcorrel}
\end{equation}%
the (field-dependent) damping parameter $\gamma _{GL}\left( H\right) \equiv 
\widetilde{\gamma }_{GL}\left( H\right) \pi \alpha /8$ should be determined
self consistently with the life time of the fluctuation modes. To determine
the TDGL propagator $\widehat{L}$ it would be convenient to go to the
wavenumber-frequency representation and exploit its relation (Eq.\ref%
{NormalizRelat}) to the microscopically derived dynamical fluctuation
propagator (see its static version, Eq.\ref{D(q)^H}):

\begin{eqnarray}
&\mathcal{D}^{H}\left( q,\Omega _{k}\rightarrow -i\Omega \right) \simeq 
\frac{1}{N_{2D}}\frac{1}{\varepsilon _{H}+\eta \left( H\right) \hbar \frac{%
Dq^{2}-i\Omega }{4\pi k_{B}T}}  \notag \\
&=\frac{\alpha k_{B}T}{N_{2D}}L\left( q,\Omega \right)  \label{D(q,-iw)^H}
\end{eqnarray}

Now, exploiting Eq.\ref{LFcorrel} in evaluating the correlation function,
that is:%
\begin{equation*}
\left \langle \phi ^{\ast }\left( \mathbf{q};t\right) \phi \left( \mathbf{q}%
;0\right) \right \rangle _{H}=2k_{B}T\hbar \gamma _{GL}\left( H\right) \int 
\frac{d\Omega }{2\pi }e^{-i\Omega t}\left \vert L\left( q,\Omega \right)
\right \vert ^{2}
\end{equation*}%
and using Eq. \ref{NormalizRelat} together with Eq.\ref{D(q,-iw)^H}, the
frequency integration is easily performed by the residue method, to find:%
\begin{eqnarray}
&&\left \langle \phi ^{\ast }\left( \mathbf{q};t\right) \phi \left( \mathbf{q%
};0\right) \right \rangle _{H}=\frac{\widetilde{\gamma }_{GL}\left( H\right) 
}{\widetilde{\eta }\left( H\right) }\left( \frac{1}{\alpha }\frac{1}{%
\varepsilon _{H}+\eta \left( H\right) \frac{\hbar Dq^{2}}{4\pi k_{B}T}}%
\right)  \notag \\
&&\times \exp \left[ -\frac{\varepsilon \left( q;H\right) }{\widetilde{\eta }%
\left( H\right) \gamma _{GL}}t/\hbar \right]  \label{phiphi}
\end{eqnarray}%
where the energy of the $q$ fluctuation-mode is given by:

\begin{equation}
\varepsilon \left( q;H\right) =k_{B}T\alpha \left( \widetilde{\varepsilon }%
_{H}+\frac{\eta \left( H\right) \hbar }{4\pi k_{B}T}Dq^{2}\right)
\label{eps(q:H)}
\end{equation}

At this point one note that in order that the (field-dependent) damping
parameter, $\gamma _{GL}\left( H\right) =\widetilde{\gamma }_{GL}\left(
H\right) \gamma _{GL}$, of the Langevin force correlator (Eq.\ref{LFcorrel}%
), be determined consistently with the characteristic rate of damping of the
correlation function in Eq.\ref{phiphi} $\widetilde{\gamma }_{GL}\left(
H\right) $ should satisfy the identity:

\begin{equation}
\widetilde{\gamma }_{GL}\left( H\right) =\widetilde{\eta }\left( H\right)
\label{consistCond}
\end{equation}%
which reduces Eq.\ref{phiphi} to the equivalent of the
fluctuation-dissipation theorem, that is:

\begin{equation}
\left \langle \phi ^{\ast }\left( \mathbf{q};t\right) \phi \left( \mathbf{q}%
;0\right) \right \rangle _{H}=\left \langle \left \vert \phi \left( \mathbf{q%
}\right) \right \vert ^{2}\right \rangle _{H}\exp \left( -\frac{t}{\tau
_{GL}\left( q;H\right) }\right)  \label{FluctDiss}
\end{equation}%
where 
\begin{equation}
\tau _{GL}\left( q;H\right) \equiv \hbar \frac{\widetilde{\eta }\left(
H\right) \gamma _{GL}}{\varepsilon \left( q;H\right) }=\hbar \frac{\gamma
_{GL}\left( H\right) }{\varepsilon \left( q;H\right) }  \label{tau_GL(q;H)}
\end{equation}%
is the life-time of the fluctuation mode at wavelength $q$. Evidently, as
seen in Eq.\ref{FluctDiss}, the auto-correlation function, $\left \langle
\phi ^{\ast }\left( \mathbf{q};0\right) \phi \left( \mathbf{q};0\right)
\right \rangle _{H}$ is found to be equal to the equilibrium momentum
distribution function, $\left \langle \left \vert \phi \left( \mathbf{q}%
\right) \right \vert ^{2}\right \rangle _{H}$ , given by Eq.\ref{momdistrH}.

Returning to the microscopic evaluation of the DOS conductivity, the field
dependence, as well as the $q$ dependence, of the Cooperon factor, has
different physical origin than that of the remaining integrand factor $%
\left
\langle \left \vert \phi \left( \mathbf{q}\right) \right \vert
^{2}\right
\rangle _{H}$ \ in Eq.\ref{sig_xx^(5+7)(H)}, which appears as a
proper finite-field extension of the zero-field CPFs momentum distribution,
Eq.\ref{momentdistrH0}. Under these circumstances the definition of a
field-dependent relaxation time for impurity scattering of electrons, $\tau
_{SO}\left( q;H\right) $, in Eq.\ref{tau_SO(q;H)} is a natural finite-field
extension of Eq.\ref{tau_SO(q)} for $\tau _{SO}\left( q\right) $ at zero
field.

This clear separability between the (collective) fluctuation effect and the
single-electron effect in the microscopic theory enables us to define an
overall CPFs density:%
\begin{eqnarray}
&&n_{CPF}\left( H\right) \equiv \frac{1}{d}\int \frac{d^{2}q}{\left( 2\pi
\right) ^{2}}\left \langle \left \vert \phi \left( \mathbf{q}\right) \right
\vert ^{2}\right \rangle _{H}  \notag \\
&=&\left( \frac{7\zeta \left( 3\right) E_{F}}{4\pi ^{2}k_{B}T}\right) \frac{%
\pi }{d}\int_{0}^{q_{c}^{2}}\frac{d\left( q^{2}\right) }{\left( 2\pi \right)
^{2}}\frac{1}{\varepsilon _{H}+\xi ^{2}\left( H\right) q^{2}}  \label{n_S(H)}
\end{eqnarray}%
even in the general case where $\tau _{SO}\left( q;H\right) $ depends on $q$
(see Eq.\ref{tau_SO(q;H)}). Here the field-dependent CPF coherence length is
given by: 
\begin{equation}
\xi \left( H\right) =\sqrt{\widetilde{\eta }\left( H\right) \frac{\pi \hbar D%
}{8k_{B}T}}  \label{xi(H)}
\end{equation}

The high-field case of interest to us allows, however, to neglect the
relatively weak $q$ dependence in Eq.\ref{tau_SO(q;H)}, so that the DOS
conductivity at finite field:%
\begin{equation}
\sigma _{DOS}^{mic}\left( H\right) =2\sigma _{xx}^{LV\left( 5+7\right)
}\left( H\right)  \label{sig_DOS=2sig_xx}
\end{equation}%
takes the simple Drude-like form:

\begin{equation}
\sigma _{xx}^{LV\left( 5+7\right) }\left( H\right) \simeq -2\frac{e^{2}}{%
m^{\ast }}n_{S}\left( H\right) \tau _{SO}\left( H\right)
\label{sig_DOS-HDrude}
\end{equation}%
with:%
\begin{equation}
\tau _{SO}\left( H\right) \equiv \widetilde{\tau }_{SO}\left( H\right) \tau
_{SO}\equiv \frac{\psi ^{\prime \prime }\left( \frac{1}{2}+\frac{\left( \mu
_{B}H\right) ^{2}}{2\pi k_{B}T\varepsilon _{SO}}\right) }{\psi ^{\prime
\prime }\left( \frac{1}{2}\right) }\tau _{SO}  \label{tau_SO(H)}
\end{equation}

Performing the integration in Eq.\ref{n_S(H)} up to the cut off at $%
q_{c}^{2}=\xi ^{-2}\left( H\right) $ (see Appendix E for further discussion
of the cutoff), and substituting in Eq.\ref{sig_DOS-HDrude}, it is written
in terms of the basic dimensionless field-dependent parameters $\widetilde{%
\eta }\left( H\right) $ and $\widetilde{\tau }_{SO}\left( H\right) $, so
that: 
\begin{equation}
\sigma _{DOS}^{mic}\left( H\right) =-\frac{e^{2}}{d\hbar }\left( \frac{%
14\zeta \left( 3\right) }{\pi ^{4}}\right) \frac{\widetilde{\tau }%
_{SO}\left( H\right) }{\widetilde{\eta }\left( H\right) }\ln \left( 1+\frac{1%
}{\varepsilon _{H}}\right)  \label{sig_DOS(H)^mic}
\end{equation}

The key parameter in Eq.\ref{n_S(H)}, through Eq.\ref{xi(H)}, is the
normalized reduced stiffness function $\widetilde{\eta }\left( H\right) $,
defined in Eq.\ref{tilde(eta(H))} (see Eq.\ref{eta(H)}), which under the
dirty limit conditions, according to Eq.\ref{eta(H)approx}, can be rewritten
in the form: 
\begin{equation}
\eta \left( H\right) \approx \psi ^{\prime }\left[ \frac{1}{2}\left( 1+\frac{%
T_{H}}{T}\right) \right]  \label{eta(TH/T)}
\end{equation}%
where the characteristic temperature $T_{H}$ is defined as:%
\begin{equation}
T_{H}\equiv \frac{\left( \mu _{B}H\right) ^{2}}{\pi k_{B}\varepsilon _{SO}}
\label{T_H}
\end{equation}

\begin{center}
\begin{figure*}[tbh]
\includegraphics[width=.4\textwidth]{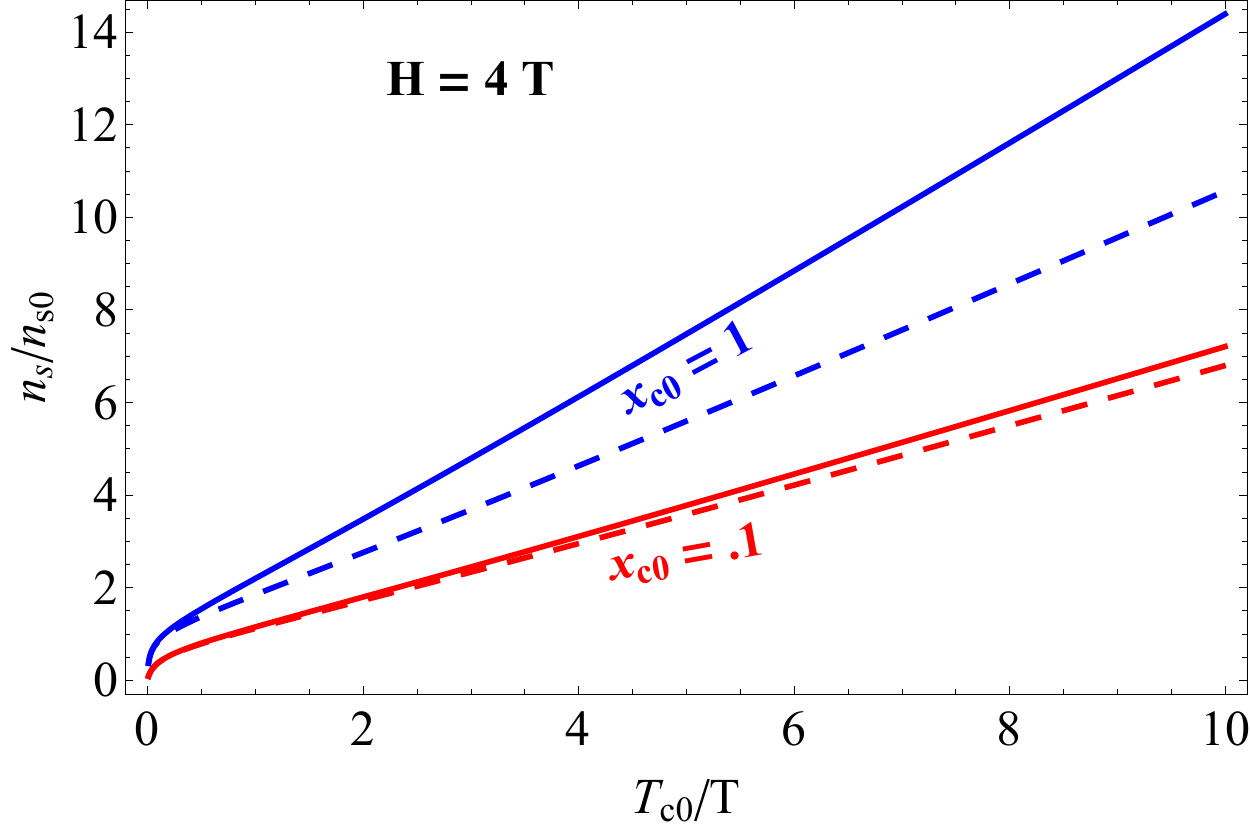} \llap{\parbox[b]{4.6in}{(a)%
\\ \rule{0ex}{1.6in} }} \includegraphics[width=.4\textwidth]{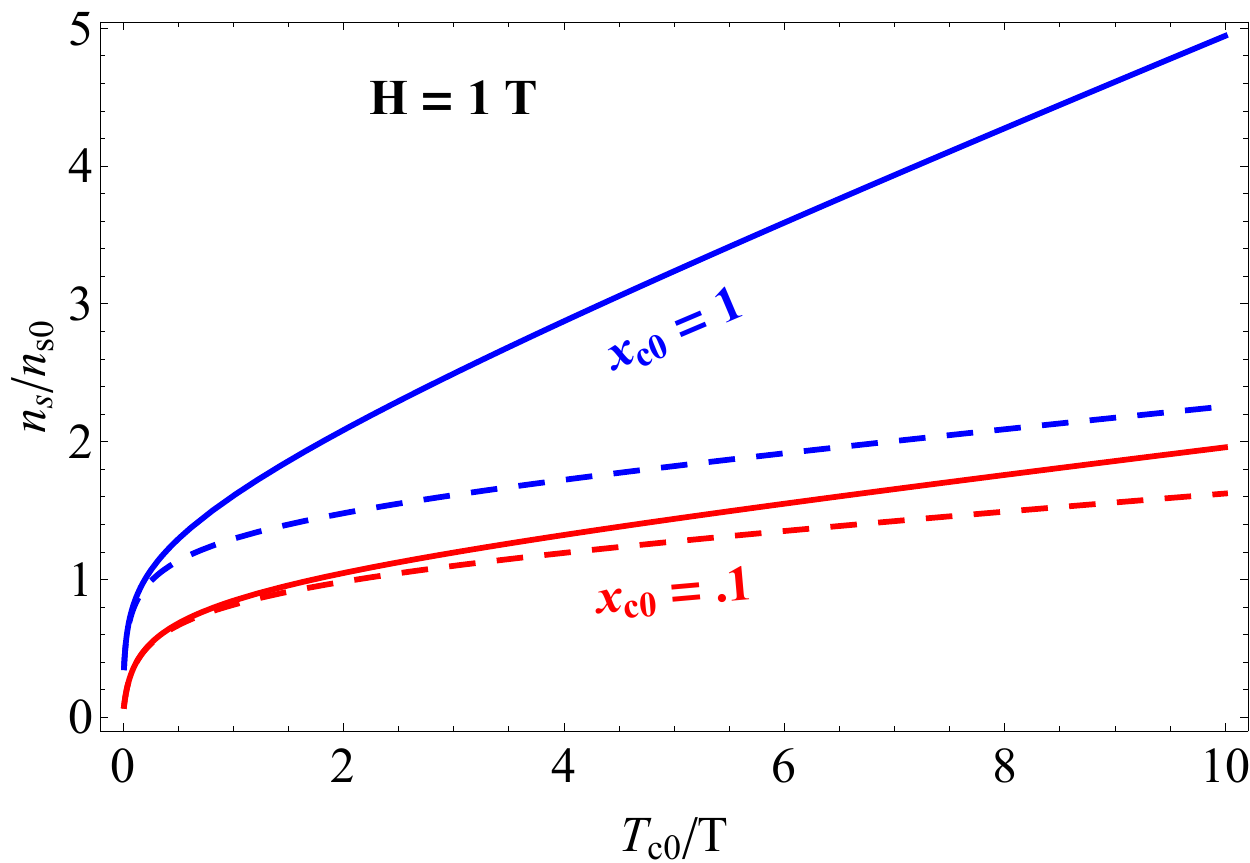} 
\llap{\parbox[b]{4.6in}{(b)\\ \rule{0ex}{1.6in}
}} \newline
\includegraphics[width=.4\textwidth]{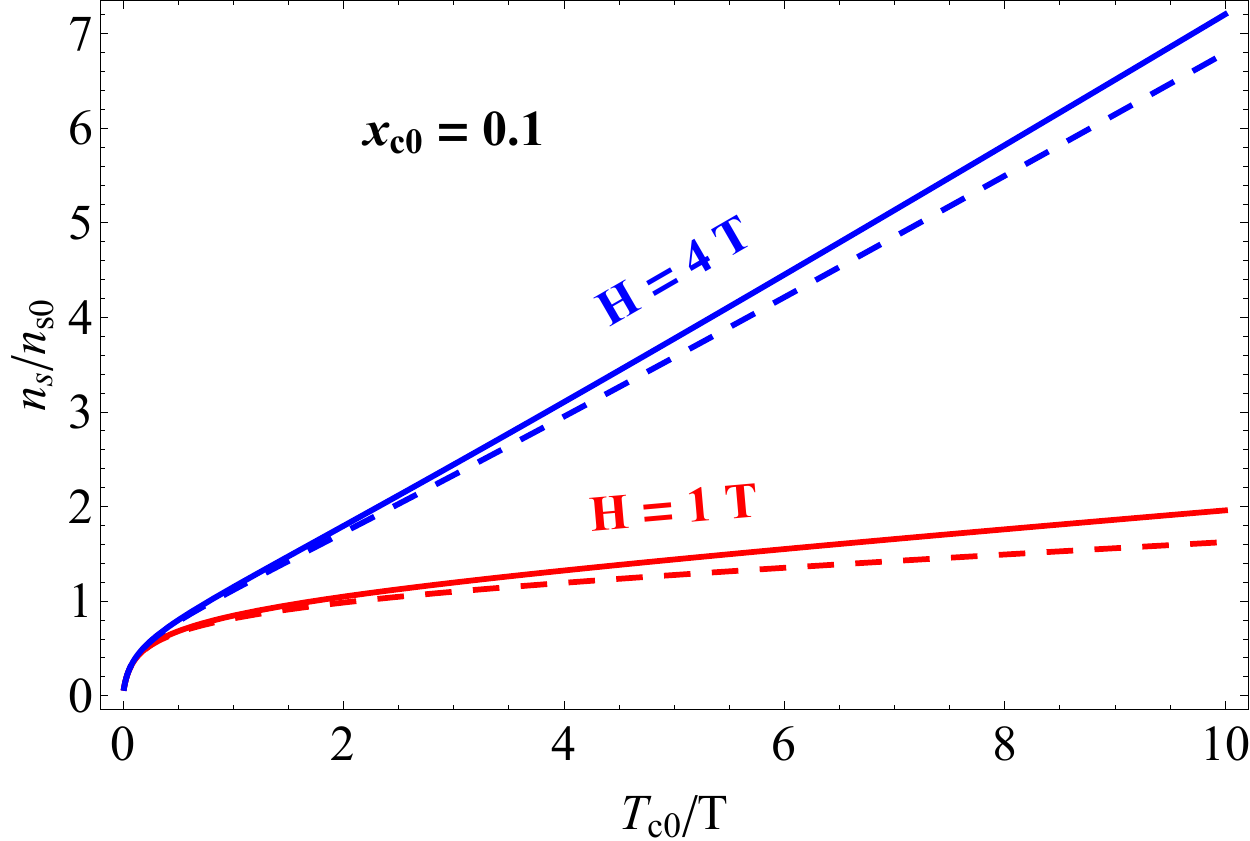} 
\llap{\parbox[c]{.6in}{(c)\\
\rule{0ex}{.7in} }}
\caption{The CPFs density $n_{S}\left( H\right) $ (solid lines) and its
approximation $n_{S}^{L}\left( H\right) $ (dashed lines) as functions of $%
T_{c0}/T$ (see Appendix E) for two characteristic values of the magnetic
field $H$ and the cutoff parameter $x_{c0}$. (a): $H=4\boldsymbol{\mathit{T}}
$; $x_{c0}=1$ (blue), $x_{c0}=0.1$ (red). (b): $H=1\boldsymbol{\mathit{T}}$
; $x_{c0}=1$, (blue), $x_{c0}=0.1$ (red).(c): $x_{c0}=1$; $H=4\boldsymbol{\ 
\mathit{T}}$ (blue), $H=1\boldsymbol{\mathit{T}}$ (red). In all graphs $%
\widetilde{\protect\varepsilon }_{H}$ was treated as independent parameter
with the value $\widetilde{\protect\varepsilon }_{H}=0.01$. The selected
values of the basic other parameters: $T_{c0}=212$m$\boldsymbol{\mathit{K}}$
and $\protect\varepsilon _{SO}=3\times 10^{-3}\boldsymbol{\mathit{eV}}$,
yield (see Eq.\protect\ref{T_H}): $T_{H=4\boldsymbol{\mathit{T}}}=33$m$%
\boldsymbol{\mathit{K}}$. Note the asymptotic linear dependence on $T_{c0}/T$
of all graphs above certain values of $T_{c0}/T$, which depend on the values
of the field $H$ and the cutoff parameter $x_{c0}$.}
\label{fig3}
\end{figure*}
\end{center}

At low temperatures and sufficiently high fields, where $T_{H}/T\gg 1$, the
asymptotic form of the digamma function yields vanishing stiffness at any
finite field, i.e.:

\begin{equation}
\eta \left( H\right) \rightarrow \frac{2T}{T_{H}},T\ll T_{H}
\label{eta(H)T0}
\end{equation}%
which remains finite, however, (equal to $\eta \left( 0\right) =\pi ^{2}/2$)
at zero field.

This extreme low temperature softening of the fluctuation modes at finite
field results in divergent CPFs density (see Fig.3 and more details in
Appendix E):%
\begin{equation}
n_{CPF}\left( H\right) \rightarrow n_{CPF}^{0}\left( \frac{T_{H}}{T}\right)
\ln \left( 1+\frac{1}{\varepsilon _{H}}\right) ,T\ll T_{H}  \label{n_S(H)T0}
\end{equation}%
where the field and temperature independent CPFs density parameter $%
n_{S}^{0} $ is defined as:%
\begin{equation}
n_{CPF}^{0}\equiv \left( \frac{7\zeta \left( 3\right) E_{F}}{4\pi ^{2}\hbar D%
}\right) \frac{1}{d}=\frac{7\zeta \left( 3\right) }{2\pi }\frac{1}{d}%
N_{2D}\varepsilon _{SO}  \label{n_s0}
\end{equation}

This divergence is, of course unphysical, since $n_{CPF}\left( H\right) $
could not exceed twice the normal-state electron density $%
n_{0}=k_{F}^{2}/2\pi $, reflecting the neglect of quantum tunneling of CPFs
in our model at zero temperature (see more details in Refs. \cite{MZPRB2021},%
\cite{MZJPC2023} and in the discussion section below).

The corresponding limiting coherence length is finite, diminishing with
increasing magnetic field: 
\begin{equation}
\xi \left( H\right) \rightarrow \frac{1}{\mu _{B}H}\sqrt{\varepsilon
_{SO}\hbar D/2}=\frac{\hbar v_{F}}{2\mu _{B}H},T\ll T_{H}  \label{xi(H)T0}
\end{equation}

In contrast, at zero field, the reduced stiffness function remains finite at
any temperature (i.e. $\widetilde{\eta }\left( H=0\right) =1$), and the
resulting CPFs density is also finite, equal to:

\begin{equation}
n_{CPF}\left( H=0\right) =n_{CPF}^{0}\left( \frac{2}{\pi ^{2}}\right) \ln
\left( 1+\frac{1}{\varepsilon }\right)  \label{n_S(H=0)}
\end{equation}%
but with infinitely long coherence length for $T\rightarrow 0$:%
\begin{equation}
\xi \left( H=0\right) \rightarrow \frac{1}{4}\hbar v_{F}\sqrt{\frac{\pi }{%
k_{B}T\varepsilon _{SO}}},T\ll T_{H}  \label{xi(H=0)T0}
\end{equation}

It will be instructive at this point to use in Eq.\ref{n_S(H)} an extension
of Eq.\ref{momdistrH} for the momentum distribution function to large
wavenumbers, which can be derived in the dirty limit, $\hbar /\tau
_{SO}=\varepsilon _{SO}>>\mu _{B}H,k_{B}T$, i.e. (see Appendix E):

\begin{equation}
\left \langle \left \vert \phi \left( \mathbf{q}\right) \right \vert
^{2}\right \rangle _{H}=\frac{7\zeta \left( 3\right) E_{F}}{4\pi ^{2}k_{B}T}%
\frac{1}{\Phi \left( q;H\right) }  \label{momdistrHex}
\end{equation}%
with: 
\begin{eqnarray}
\Phi \left( q;H\right) &=&\widetilde{\varepsilon }_{H}+\psi \left[ \frac{1}{2}%
\left( 1+\frac{T_{H}}{T}\right) +\frac{T_{c0}}{T}\frac{2}{\pi ^{2}}\xi
_{0}^{2}q^{2}\right] \notag \\
&-&\psi \left[ \frac{1}{2}\left( 1+\frac{T_{H}}{T}\right) %
\right] \label{Phi(q;H)}
\end{eqnarray}
and (see Eq.\ref{xi,gam_GL}): 
\begin{equation}
\xi _{0}\equiv \xi \left( T_{c0}\right) =\sqrt{\frac{\hbar \pi D}{%
8k_{B}T_{c0}}}  \label{xi_0}
\end{equation}

Note the decorated notation $\widetilde{\varepsilon }_{H}$ used explicitly
in Eq.\ref{Phi(q;H)} for the positive-definite critical shift parameter,
which was introduced originally in Ref. \cite{MZPRB2021} to take into
account, self-consistently, the effect of interaction between fluctuations
on $\varepsilon _{H}$ defined in Eq.\ref{eps_H} for free (Gaussian)
fluctuations. We shall use this notation for the critical shift parameter
from now on in this paper to remind the reader that it can have only
nonvanishing positive values, except for $T=0$, consistently with the
absence of zero resistance in the experimental data.

The importance of the vanishing reduced stiffness function, $\widetilde{\eta 
}\left( H\right) \simeq \left( 2/\pi ^{2}\right) \psi ^{\prime }\left(
1/2+T_{H}/2T\right) \rightarrow \left( 2/\pi ^{2}\right) \left(
2T/T_{H}\right) $, of the fluctuation modes at low temperatures, $\left(
T/T_{H}\right) \ll 1$, becomes transparent only after expanding the energy
function, Eq.\ref{Phi(q;H)}, to first order in the kinetic energy term,
which yields in agreement with Eq.\ref{momdistrH}: $\Phi \left( q;H\right)
\simeq \widetilde{\varepsilon }_{H}+\widetilde{\eta }\left( H\right) \left(
T_{c0}/T\right) \xi _{0}^{2}q^{2}=\widetilde{\varepsilon }_{H}+\widetilde{%
\eta }\left( H\right) \left( \eta _{\left( 2\right) }q^{2}\right) $ (see
Appendix E). Continuing analytically the Taylor expansion of $\Phi \left(
q;H\right) $ in $\widetilde{\eta }\left( H\right) \left( T_{c0}/T\right) \xi
_{0}^{2}q^{2}\rightarrow \left( 2/\pi ^{2}\right) \left(
2T_{c0}/T_{H}\right) \xi _{0}^{2}q^{2}$, it takes the low temperature form
(see Appendix E): 
\begin{equation}
\Phi \left( q;H\right) \rightarrow \widetilde{\varepsilon }_{H}+\ln \left[
1+\left( \frac{2}{\pi ^{2}}\right) \left( \frac{2T_{c0}}{T_{H}}\right) \xi
_{0}^{2}q^{2}\right]  \label{Phi(q;H)asymp}
\end{equation}%
in which the logarithmic term does not depend on temperature ! The condition
for the validity of the linear approximation, uniformly for all $q$ values
up to the cutoff $q_{c}$: 
\begin{equation}
\left( \frac{T_{c0}}{T_{H}}\right) x_{c0}\ll 1  \label{linapprox}
\end{equation}%
where: 
\begin{equation}
x_{c0}\equiv \frac{\hbar Dq_{c}^{2}}{4\pi k_{B}T_{c0}}=\frac{2}{\pi ^{2}}\xi
_{0}^{2}q_{c}^{2}  \label{xc0}
\end{equation}%
depends both on the field, through $T_{H}$, and on the cutoff wavenumber,
through $x_{c0}$ (see Fig.3), but does not depend on temperature.

It is remarkable that even for larger values of the cutoff parameter $x_{c0}$%
, where the linear approximation (see Eq.\ref{linapprox}) is not valid, the
low temperature (asymptotic) CPFs density, given by: 
\begin{equation}
n_{CPF}\left( H\right) \rightarrow n_{CPF}^{0}\left( \frac{T_{H}}{T}\right)
\int_{0}^{x_{c0}T_{c0}/T_{H}}\frac{d\chi }{\widetilde{\varepsilon }_{H}+\ln
\left( 1+\chi \right) }  \label{n_SLT}
\end{equation}%
has the same $1/T$ divergence at $T\ll T_{H}$ as the linear approximation
given by Eq.\ref{n_S(H)T0}, but with a larger slope, as shown in Fig.3 (see
also Appendix E). It should be emphasized that this divergence is not
peculiar to the strong spin-orbit scattering situation considered in the
present paper. As indicated in Sec.IV, the field-induced ultimate softening
of the fluctuation modes responsible for this divergence takes place at
sufficiently low temperature at any field that induces Zeeman spin-splitting
(pair-breaking) frequency, $2\mu _{B}H/\hbar $, smaller than the
(pair-compensating) spin-orbit relaxation rate $1/\tau _{SO}$.

\subsection{The crossover to localization of Cooper-pair fluctuations}

The discontinuous nature of the reduced stiffness function $\eta \left(
H\right) $ at $H=0$ in the zero-temperature limit, revealed in Eq. \ref%
{eta(H)T0}, is inherent to peculiar features of the CPFs at low
temperatures. These features may be best appreciated by considering the
Fourier transform to real space of the momentum distribution function, Eq.%
\ref{momdistrH}, in the low temperature limit $T\ll T_{H}$, where
field-induced condensation in real space is expected on the basis of the
limiting behavior of $n_{S}\left( H\right) $ and $\xi \left( H\right) $
according to Eqs.\ref{n_S(H)T0} and \ref{xi(H)T0}, respectively. In the
absence of interaction between fluctuations this Fourier transform can be
shown to be related, by Eq.\ref{FluctDiss}, to the equal-time real-space
Cooper-pair amplitude correlation function through the equation:

\begin{eqnarray}
&\left \langle \phi ^{\ast }\left( \mathbf{r+\rho ,}t_{0}\right) \phi \left( 
\mathbf{r,}t_{0}\right) \right \rangle _{H}=\left( \frac{1}{2\pi }\right)
^{2}\int d^{2}q\left \langle \left \vert \phi \left( \mathbf{q}\right)
\right \vert ^{2}\right \rangle _{H}e^{i\mathbf{q\cdot \rho }}  \notag \\
&\equiv \left \langle \left \vert \phi \left( \mathbf{\rho }\right) \right
\vert ^{2}\right \rangle _{H}  \label{CPampCorr}
\end{eqnarray}

The last identity indicates that, as a function of $\mathbf{\rho }$, this
correlation function is a real space measure of the CPFs density, which
determines the probability amplitude for a CPF, generated at any point $%
\mathbf{r}$ of the underlying uniform 2D system, to propagate a distance $%
\rho =\left \vert \mathbf{\rho }\right \vert $.

Performing the angular integration in Eq.\ref{CPampCorr} and using the more
general expression for the momentum distribution function in the dirty
limit, Eq.\ref{momdistrHex}, i.e. with $\Phi \left( q;H\right) $ given by Eq.%
\ref{Phi(q;H)}, it can be rewritten in the form:%
\begin{equation}
\left \langle \left \vert \phi \left( \mathbf{\rho }\right) \right \vert
^{2}\right \rangle _{H}=\frac{1}{2\pi }n_{S}^{0}\xi _{0}^{-2}g\left( \rho
/\xi _{0}\right)  \label{corrfunct}
\end{equation}%
where $g\left( \rho /\xi _{0}\right) $ is given by:

\begin{equation}
g\left( \rho /\xi _{0}\right) =\frac{T_{c0}}{T}\int_{0}^{\widetilde{q}_{c}}%
\frac{\widetilde{q}J_{0}\left( \widetilde{q}\rho /\xi _{0}\right) }{\Phi
\left( \widetilde{q};H\right) }d\widetilde{q}  \label{g(rho)}
\end{equation}%
$\widetilde{q}\equiv \xi _{0}q$, and $J_{0}\left( \widetilde{q}\rho /\xi
_{0}\right) $ is the zero order Bessel function of the first kind.

The dependence of the dimensionless density function $g\left( \rho /\xi
_{0}\right) $ on $\rho $ has a decaying envelope, modulated by an
oscillatory function associated with the sharp cutoff $q_{c}$. The length
scale of this attenuation is the localization length, given by (see Appendix
G):

\begin{equation}
\rho _{loc}\left( H\right) =\left( \frac{T_{c0}}{T}\frac{\widetilde{\eta }%
\left( H\right) }{\widetilde{\varepsilon }_{H}}\right) ^{1/2}\xi _{0}
\label{rho_loc}
\end{equation}

The results of a detailed analysis of $g\left( \rho /\xi _{0}\right) $ and
its asymptotic behavior (see Appendix G) are shown in Fig.4, where the
condensing CPFs at diminishing temperature are found to pile up within a
region of diminishing size of the order of $\rho _{loc}\left( H\right) $
under increasing field $H$.

The positive-definite self-consistent critical shift parameter, $\widetilde{%
\varepsilon }_{H}$, appearing in Eq.\ref{rho_loc}, which is a monotonically
increasing function of the field $H$, is used in the calculations as an
independent free parameter, in order to illustrate how the minimal gap, $%
\widetilde{\varepsilon }_{H}$, of the energy spectrum given in Eq.\ref%
{Phi(q;H)}, influences the condensation and localization of CPFs in real
space.

\begin{center}
\begin{figure*}[tbh]
\includegraphics[width=.45\textwidth]{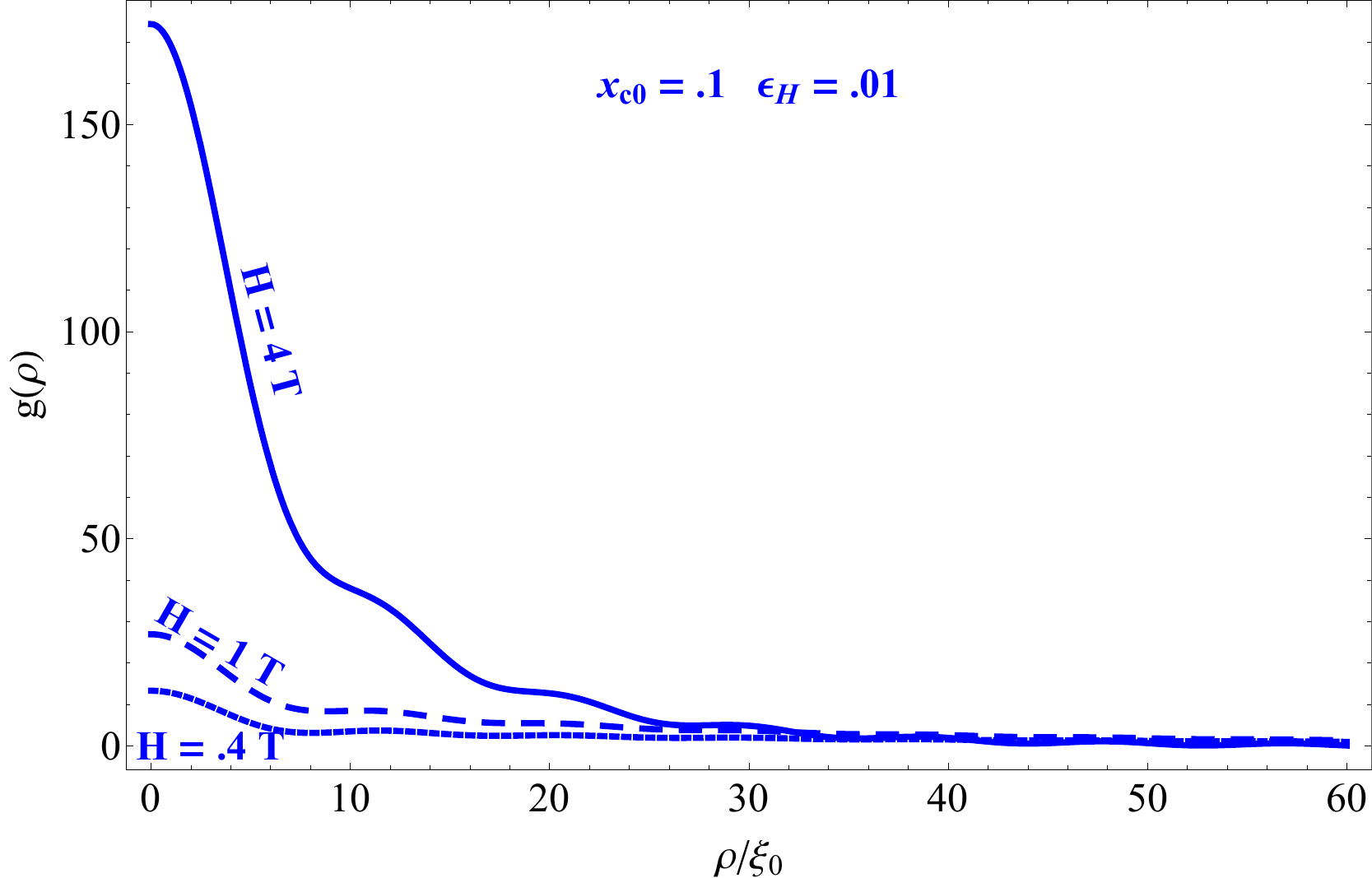} \llap{\parbox[b]{.7in}{(a)%
\\ \rule{0ex}{1.6in} }} \includegraphics[width=.45\textwidth]{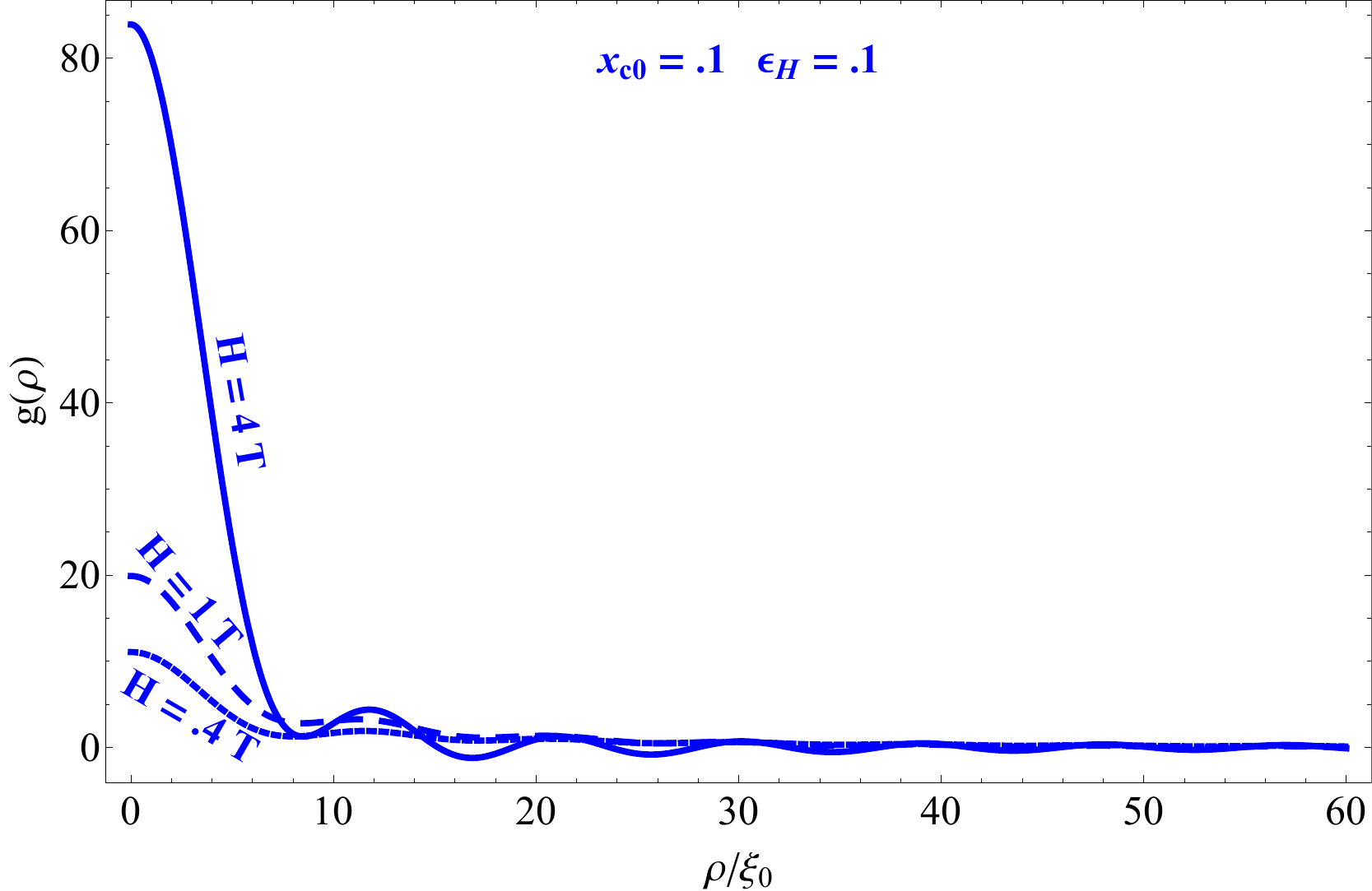} %
\llap{\parbox[b]{.7in}{(b)\\ \rule{0ex}{1.6in} }} \newline
\includegraphics[width=.45\textwidth]{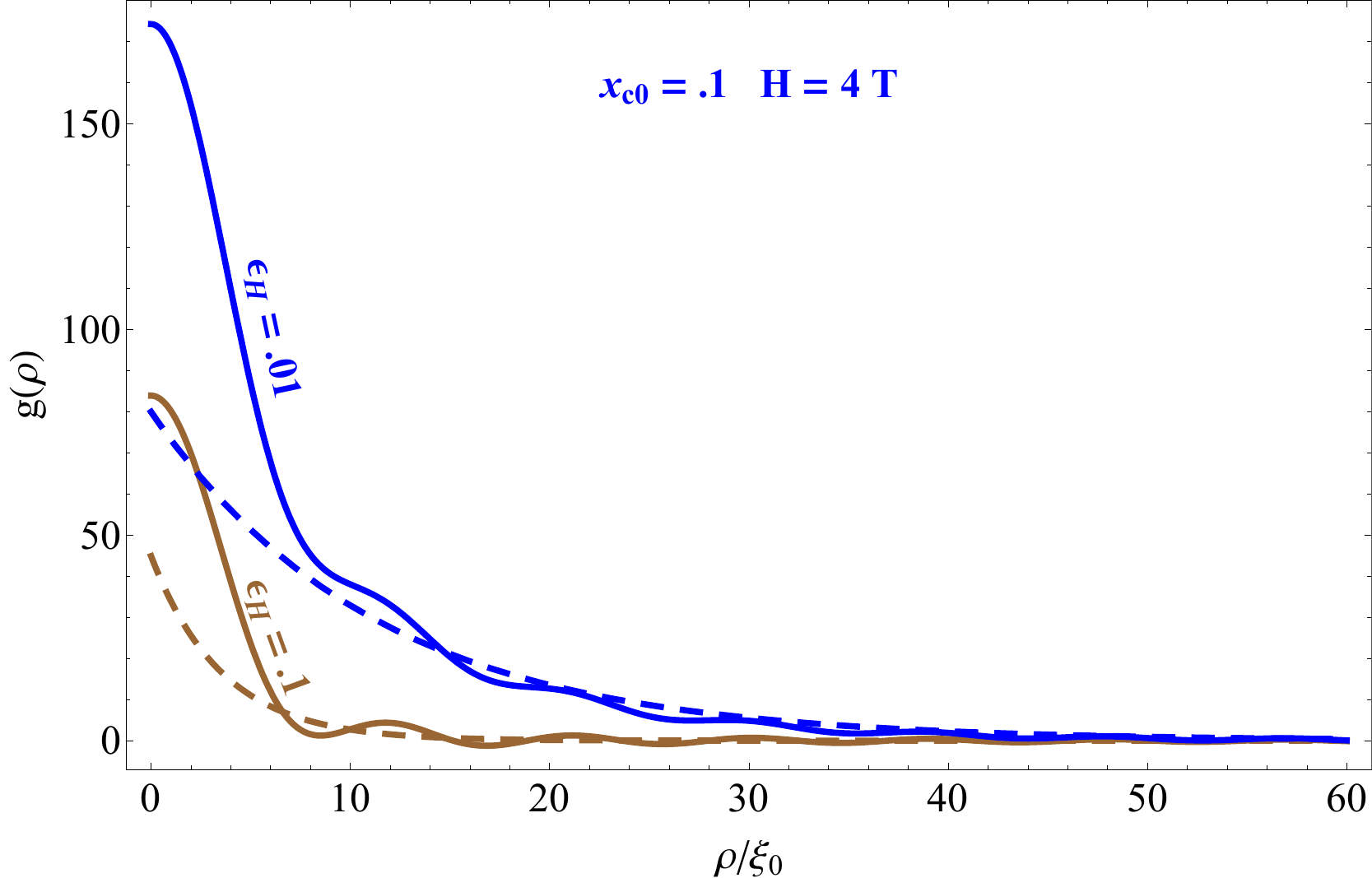} \llap{\parbox[b]{.7in}{(c)%
\\ \rule{0ex}{1.6in} }} \includegraphics[width=.45\textwidth]{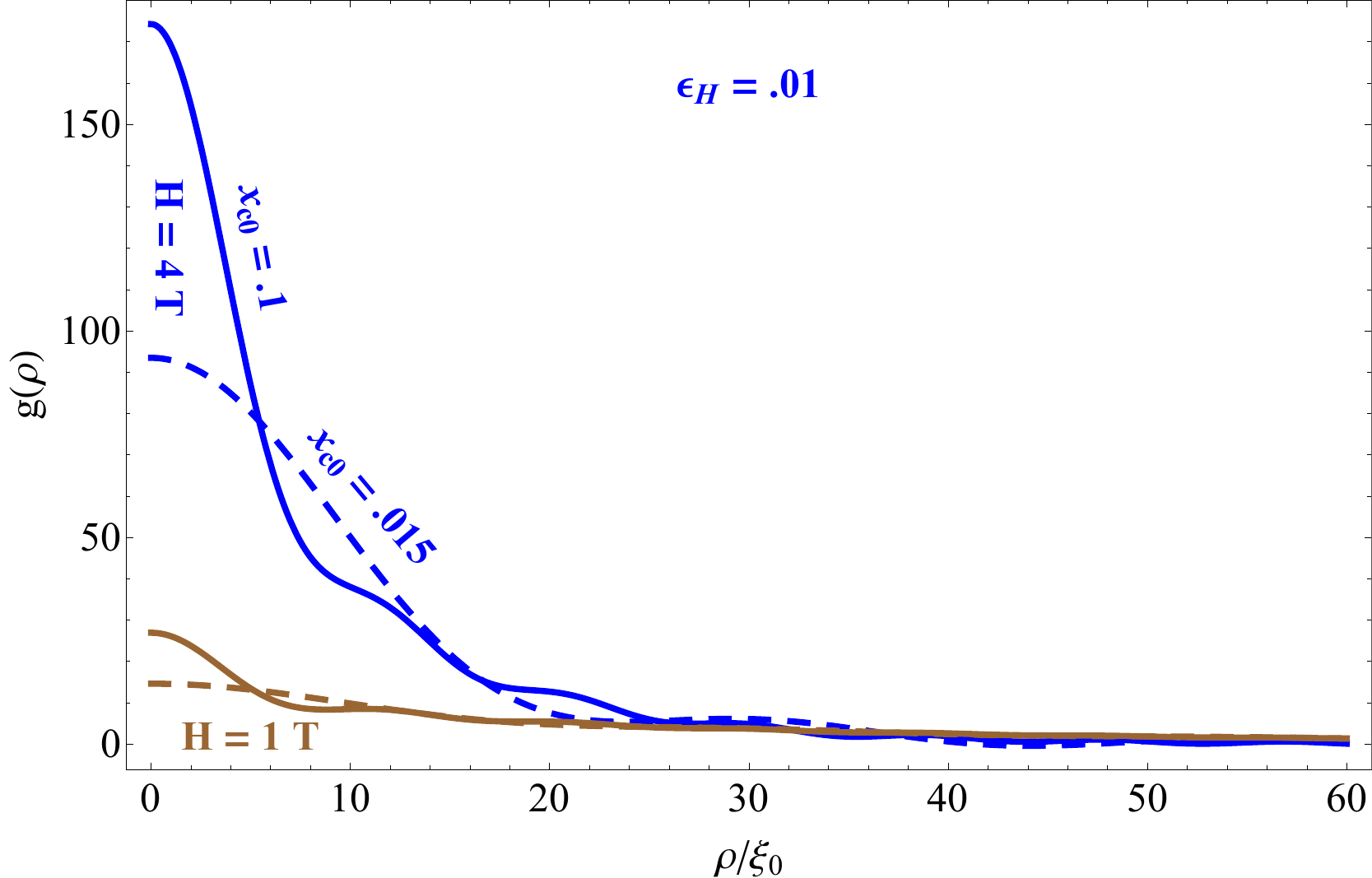} %
\llap{\parbox[b]{.7in}{(d)\\ \rule{0ex}{1.6in} }}
\caption{Dimensionless real-space CPFs density $g\left( \protect\rho /%
\protect\xi _{0}\right) $, illustrating the phenomena of condensation and
localization of CPFs, as discussed in the text, calculated at temperature $%
T=2$mK for different values of the magnetic field, $H$ , the cutoff
parameter $x_{c0}$, and $\widetilde{\protect\varepsilon }_{H}$, which is
considered for illustration as independent free parameter. (a) $x_{c0}=.1$; $%
\widetilde{\protect\varepsilon }_{H}=.01$; $H=4$T,$1$T,$.1$T. (b) $x_{c0}=.1$
;$\widetilde{\protect\varepsilon }_{H}=.1$ ; $H=4$T,$1$T,$.1$T. (c) $%
x_{c0}=.1$; $H=4$T ; $\widetilde{\protect\varepsilon }_{H}=.01$(blue), $%
\widetilde{\protect\varepsilon }_{H}=.1$(brown), exponential asymptotes
(dashed lines) (see Appendix G). (d) $\widetilde{\protect\varepsilon }%
_{H}=.01$; $x_{c0}=.1$ (solid lines), $x_{c0}=.015$ (dashed lines); $H=4T$
(blue), $H=1T$ (brown).}
\label{fig4}
\end{figure*}
\end{center}

The CPFs density function defined in Eq.\ref{corrfunct}, which shows
localization of fluctuations propagated around any arbitrary point in the 2D
system, has been derived with the assumption that interactions between
fluctuations are neglected. Thus, it is expected that Coulomb repulsion
between the fluctuations would lead to formation of an extended 2D structure
of mesoscopic puddles. This is qualitatively plausible conclusion, though we
have not gone beyond that point quantitatively since the details of this
structure is not relevant to our purposes here (see the remark at the end of
Sec.VI).

One should also note that the validity of Eq.\ref{corrfunct} as a local
density of some complex bosonic particles as a function of the distance from
their point of creation is limited by the life-time $\tau _{GL}\left(
q;H\right) $ of the corresponding excitation, as given by Eq. \ref%
{tau_GL(q;H)}. At low temperatures, $T\ll T_{H}$, where:

\begin{equation*}
\tau _{GL}\left( q;H\right) \rightarrow \frac{\hbar /4\pi k_{B}T_{H}}{%
\widetilde{\varepsilon }_{H}+\hbar Dq^{2}/4\pi k_{B}T_{H}}
\end{equation*}%
our estimate of this characteristic time at $H=4T$, and for the typical
experimental parameters (for which $T_{H}=33mK$, and the cutoff parameter $%
x_{c0}\approx 0.015$), is: $\tau _{GL}\left( q_{c};H=4T\right) \approx
10^{-10}\boldsymbol{\mathit{s}}$, that is about three orders of magnitude
larger than the typical electron relaxation time $\tau _{SO}=\hbar
/\varepsilon _{SO}\approx 10^{-13}\boldsymbol{\mathit{s}}$.

To summarize, upon increasing magnetic field, the system of CPFs crossovers
from a spatially uniform state, at zero field, to inhomogeneous states at
high fields. For sufficiently low temperatures and high fields, where $T\ll
T_{H}$, the inhomogeneous states are characterized by CPFs condensation in
mesoscopic puddles. A plausible estimate of the crossover field, $H_{cross}$%
, can be made on the scale of the reduced stiffness parameter $\widetilde{%
\eta }\left( H\right) $, half way between its edge values corresponding the
uniform and the nonuniform states, that is:

\begin{equation*}
\widetilde{\eta }\left( H_{cross}\right) =1/2
\end{equation*}%
At low temperatures, where $T\ll T_{H}$, this may be done by using Eq.\ref%
{eta(H)T0}, which yields: 
\begin{equation}
\mu _{B}H_{cross}\sim \left( \varepsilon _{SO}k_{B}T\right) ^{1/2}
\label{H_cross}
\end{equation}

For the sake of better sense of scaling, we may estimate the field, $H_{upp}$
at which the low-temperature limit of the CPFs density $n_{S}\left( H\right) 
$, Eq.\ref{n_S(H)T0}, approaches its (physical) upper limit, i.e. $1/2$ of
the total number density of electrons $n_{0}=k_{F}^{2}/2\pi d$: 
\begin{eqnarray*}
n_{CPF}\left( H\right) &\rightarrow &\left( \frac{7\zeta \left( 3\right) }{%
4\pi ^{2}}\right) \frac{1}{d}N_{2D}\frac{\left( \mu _{B}H\right) ^{2}}{k_{B}T%
}\ln \left( 1+\frac{1}{\widetilde{\varepsilon }_{H}}\right) \\
&=&\frac{1}{2}\left( \frac{1}{d}\frac{k_{F}^{2}}{2\pi }\right) =\frac{1}{2}%
n_{0}
\end{eqnarray*}%
which yields: 
\begin{equation}
\mu _{B}H_{upp}\sim \left( E_{F}k_{B}T\right) ^{1/2}  \label{H_upp}
\end{equation}

Thus, the plausible hierarchy of the parameters $E_{F}>\varepsilon _{SO}$
ensures that the crossover field, $H_{cross}$ can always be reached.

\section{Discussion and Conclusion}

The peculiar features of the system of CPFs, discussed in the previous
sections, reflect essential inconsistency of the microscopic theory of
fluctuations in superconductors at finite field and very low temperatures.
In particular, it was shown that the contribution of the basic pair of
diagrams to the DOS conductivity obtained within the conventional
microscopic approach can be expressed in terms of an effective Drude formula
(see Eq.\ref{sig_DOS-HDrude}) in which $n_{CPF}\left( H\right) $ is an
effective CPFs density (Eq.\ref{n_S(H)}) and $\tau _{SO}\left( H\right) $ is
an effective single electron relaxation time (Eq.\ref{tau_SO(H)}). These
definitions are not arbitrary; on the one hand, the field dependence of $%
\tau _{SO}\left( H\right) $ is exclusively determined by the renormalized
pairing vertex factors (see Eq.\ref{lamb^H(q,om)}), associated with electron
scatterings by background electrons via virtual exchange of CPFs. On the
other hand, the field dependence of $n_{CPF}\left( H\right) $ is associated
exclusively with the fluctuation propagator, controlled by the normalized
reduced stiffness function $\widetilde{\eta }\left( H\right) $, which tends
to zero at any $H>0$ with $T\rightarrow 0$ (see Eq.\ref{eta(H)T0}), but
remains finite (equal to $1$) at zero field. Consequently, the density $%
n_{CPF}\left( H\right) $ diverges for $T\rightarrow 0$ at any $H>0$ (see Eq.%
\ref{n_S(H)T0} and note the comment following Eq.\ref{n_s0}), but remains
finite at $H=0$ (see Eq.\ref{n_S(H=0)}). Furthermore, the divergent
Cooper-pair coherence length $\xi \left( H=0\right) $ in the $T\rightarrow 0$
limit (Eq.\ref{xi(H=0)T0}), which characterizes a homogeneous system with
long-ranged superconducting order at $H=0$, contracts at $H>0$ to a finite
localization length, where increasingly large numbers of CPFs condense under
diminishing temperature.

The microscopic theory of fluctuations in superconductors which leads to
these peculiar results is basically a perturbation theory, developed for
spatially homogenous systems, which rests upon a diagrammatic expansion of
the conductivity in the fluctuations propagator about the normal-state
conductivity. In this framework the scope of the calculations is restricted
to fluctuations effect on the normal-state conductivity above the transition
to superconductivity. Within this microscopic\ approach we may write the
total conductivity as:

\begin{equation}
\sigma _{tot}^{mic}\left( H\right) =\sigma _{n}+\sigma _{DOS}^{mic}\left(
H\right) +\sigma _{AL}^{mic}\left( H\right)  \label{sig_tot(H)^mic}
\end{equation}%
where $\sigma _{n}$ is the normal-state (zero-order in the expansion)
conductivity, $\sigma _{DOS}^{mic}\left( H\right) $ is the DOS conductivity
given by Eq.\ref{sig_DOS(H)^mic}, and $\sigma _{AL}^{mic}\left( H\right) $
is the AL conductivity written in Eq.\ref{sig_AL(H)^mic}. As indicated in
Sec.III, the Maki-Thompson conductivity (\cite{MakiPTP1968},\cite%
{ThompsonPRB1970}) is neglected here due to the presence of strong
spin-orbit scatterings (see Appendix A). Both $\sigma _{DOS}^{mic}\left(
H\right) $ and $\sigma _{AL}^{mic}\left( H\right) $ have been calculated
here within the framework of the LV method used in Ref.\cite{LV05} in which
quantum critical fluctuations \cite{Glatzetal2011},\cite{Lopatinetal05},\cite%
{Khodas2012} were neglected.

Under these circumstances at finite field $H$ and very low temperature $T\ll
T_{H}$, due to the vanishing $\widetilde{\eta }\left( H\right) \rightarrow
\left( 2/\pi ^{2}\right) T/T_{H}$ and the more quickly vanishing $\widetilde{%
\tau }_{SO}\left( H\right) \rightarrow \left( T/T_{H}\right) ^{2}/7\zeta
\left( 3\right) $, both the AL (Eq.\ref{sig_AL(H)^mic}) and DOS (Eq.\ref%
{sig_DOS(H)^mic}) conductivities:%
\begin{eqnarray}
\sigma _{AL}^{mic}\left( H\right) &\rightarrow &\frac{e^{2}}{8\pi ^{2}d\hbar 
}\left( \frac{T}{T_{H}}\right) \frac{1}{\widetilde{\varepsilon }_{H}},
\label{sig_AL(H)lim} \\
\sigma _{DOS}^{mic}\left( H\right) &\rightarrow &-\frac{e^{2}}{2\pi
^{2}d\hbar }\left( \frac{T}{T_{H}}\right) \ln \left( \frac{1}{\widetilde{%
\varepsilon }_{H}}\right)  \label{sig_DOS(H)lim}
\end{eqnarray}%
vanish with $T$. \ This linearly vanishing with temperature AL and DOS
conductivities are consistent with the results reported in Ref.\cite%
{Glatzetal2011} at very low temperatures in the absence of quantum critical
fluctuations (see a remark below Eq.\ref{Ixx^(5)}). The different field
dependencies should be related to the different magnetic field orientations
(perpendicular in Ref.\cite{Glatzetal2011} as compared to parallel in our
case).

The above discussion clearly indicates that the perturbation theory inherent
to the microscopic LV approach can not be directly applied to the extremely
inhomogeneous real-space Cooper-pairs (boson) condensate that emerges from
our analysis. Furthermore, the formal divergence of $n_{CPF}\left(
H>0\right) $ in the zero temperature limit versus the physical constraint
imposed by the conservation of the total number of electrons available for
pairing indicate that the grand canonical ensemble of electrons underlying
the microscopic theory of superconductivity is unsubstantiated here.

As discussed in detail in several recent papers (\cite{Glatzetal2011}, \cite%
{Lopatinetal05}, \cite{Khodas2012}), corrections due to quantum fluctuations
in both the DOS and the AL conductivities (as well as in Maki-Thompson
contributions, neglected here, see Appendix A) around the quantum critical
field prevent the vanishing of the fluctuation conductivity in the zero
temperature limit. The field dependence of these corrections could not
account, however, for the pronounced MR peaks observed experimentally in Ref.%
\cite{Mograbi19}.

The crucial point here is the irrelevance of the grand canonical ensemble
underlying the microscopic theory of superconductivity and the existence of
condensed mesoscopic puddles of long-lived boson excitations, which act
through pair-breaking processes as reservoirs for the remaining system of
(unpaired) normal-state electrons. Within this TDGL functional approach (see
Refs.\cite{MZPRB2021}, and \cite{MZJPC2023}) quantum tunneling of CPFs and
their pair breaking out of mesoscopic enclaves reinforce inter-puddle
transport by fermionic quasi-particle so that the conductance of the boson
excitations is exclusively represented by the Schmidt-Fulde-Maki
paraconductivity (see Eq.\ref{Q_ALgen}), whereas the residual normal-state
conductivity is due to fermionic quasi-particles. Under these circumstances
the diminishing stiffness of the fluctuation modes at low temperatures
sharply suppresses the paraconductivity against the increasing normal-state
conductivity upon increasing field \cite{DiezPRL2015}, and so leading to the
observed MR peak just above the SC critical field.

\appendix

\section{The relevant diagrams}

The literature dealing with the effect of superconducting fluctuations on
the conductivity from the point of view of the microscopic GGL theory is
quite extensive. The most comprehensive and elaborated account of this
approach can be found in Ref.\cite{LV05}. Some specific details relevant to
our analysis of the diagrams shown in Fig.1, concerning in particular, the
relative importance of the various contributions to the DOS conductivity
(diagrams 5-8 in Fig.1), can be found in earlier papers: \cite%
{AltshulerJETP1983}, \cite{Dorin-etal-PRB1993}.

A different type of diagrams (2-4 in Fig.1), corresponding to electron-hole
(Andreev-like) scatterings by CPFs \cite{ManivAlexanderJPC1976}, that is
well-known as the Maki--Thompson diagram \cite{MakiPTP1968},\cite%
{ThompsonPRB1970}, includes two parts of contributions: A singular part,
arising from the coherent Andreev-like scattering, with positive
contribution to the fluctuation conductivity, and a regular part with
negative contribution to the conductivity, similar to the DOS conductivity.
In our model we disregard this type of diagrams altogether since the strong
spin-orbit scatterings, which characterize the SrTiO$_{3}$/LaAlO$_{3}$
interfaces under consideration here, are known to destroy the coherence
responsible for the singularity \cite{LV05}, so that the remaining regular
positive contribution is cancelled, or nearly cancelled by the negative ones 
\cite{ManivAlexanderJPC1977}.

\section{The Copper-pairs (GL) current density}

Eq.\ref{j(r,t)} is obtained from the variational condition of the
electromagnetically modified GL free energy functional with respect to the
vector potential: 
\begin{equation}
\frac{\partial \mathfrak{L}\left( \Delta ,\mathbf{A}\right) }{\partial 
\mathbf{A}}+\frac{1}{8\pi }\frac{\partial }{\partial \mathbf{A}}\int
d^{3}r\left( \mathbf{\nabla }\times \mathbf{A}\right) ^{2}=0  \label{varGL_F}
\end{equation}%
in conjunction with the identity: 
\begin{equation*}
\frac{1}{8\pi }\frac{\partial }{\partial \mathbf{A}}\int d^{3}r\left( 
\mathbf{\nabla }\times \mathbf{A}\right) ^{2}=\frac{1}{4\pi }\mathbf{\nabla }%
\times \left( \mathbf{\nabla }\times \mathbf{A}\right) =\frac{1}{c}\mathbf{j}
\end{equation*}

\section{The normalization constant}

To find $\alpha $ we note that Eq.\ref{NormalizRelat}, relating the
propagators, $\mathcal{D}\left( q,\Omega \right) $ in the $\Delta $
representation to $L\left( q,\Omega \right) $ in the $\phi $ representation,
is equivalent to the normalization of the GL wavefunction:

\begin{equation}
\left \vert \phi \left( \mathbf{r}\right) \right \vert ^{2}=\frac{N_{2D}}{%
\left( \alpha k_{B}T\right) }\left \vert \Delta \left( \mathbf{r}\right)
\right \vert ^{2}  \label{normalGL}
\end{equation}

As emphasized in the main text, the normalization constant $\alpha $, in the
dirty-limit under study, may be evaluated in the clean limit, i.e. for
coherence length:%
\begin{equation}
\xi \left( T\right) \rightarrow \xi _{c}\left( T\right) =\sqrt{\frac{7\zeta
\left( 3\right) }{8}}\left( \frac{\hbar v_{F}}{2\pi k_{B}T}\right)
\label{xi(T)c}
\end{equation}%
since normalization of the wavefunctions should not depend on scatterings.
Thus, using the expression for $\alpha $ presented in Eq.\ref{alpha}, for
which the normalization takes the form:

\begin{equation}
\left \vert \phi \left( \mathbf{r}\right) \right \vert ^{2}=\frac{7\zeta
\left( 3\right) }{8}\frac{k_{F}^{2}}{2\pi }\left( \frac{\left \vert \Delta
\left( \mathbf{r}\right) \right \vert }{\pi k_{B}T}\right) ^{2}\simeq \frac{%
k_{F}^{2}}{2\pi }\left( \frac{\left \vert \Delta \left( \mathbf{r}\right)
\right \vert }{\pi k_{B}T}\right) ^{2}  \label{normalGL2}
\end{equation}%
the clean-limit GL propagator, Eq.\ref{L(q)^-1}, at zero frequency, is
written in the canonical Schrodinger-like form:

\begin{equation}
L\left( q,0\right) ^{-1}=\frac{\hbar ^{2}}{4m^{\ast }}\left[ \xi \left(
T\right) ^{-2}\varepsilon +q^{2}\right]  \label{L(q)^-1cl}
\end{equation}%
with the Cooper-pair mass equals twice the electron band mass $m^{\ast }$.

The dirty-limit momentum distribution function, Eq.\ref{momentdistrH0}, may
be therefore evaluated with $\alpha $ given by Eq.\ref{alpha} with the
dirty-limit coherence length given in Eq.\ref{xi,gam_GL}. \ The result takes
the form:

\begin{equation}
\left \langle \left \vert \phi \left( q\right) \right \vert ^{2}\right
\rangle =\frac{28\zeta \left( 3\right) }{\pi ^{2}}\left( N_{2D}\frac{\hbar }{%
\tau _{SO}}\right) \frac{1}{\xi \left( T\right) ^{-2}\varepsilon +q^{2}}
\label{momdistDirt}
\end{equation}%
with the dirty limit coherence length:%
\begin{equation}
\xi \left( T\right) \rightarrow \xi _{d}\left( T\right) =\sqrt{\frac{\pi
\hbar D}{8k_{B}T}}  \label{xi(T)d}
\end{equation}

For comparison, the clean limit result is obtained from Eq.\ref{momdistDirt}
by replacing $\hbar /\tau _{SO}$ with $\left[ 4\pi ^{3}/7\zeta \left(
3\right) \right] k_{B}T$ and $\xi _{d}\left( T\right) $ with $\xi _{c}\left(
T\right) $ for the coherence length $\xi \left( T\right) $.

Finally, the TDGL expression for the DOS conductivity, defined in terms of
the CPFs density $n_{s}$ by using using Eq.\ref{momdistDirt}, is shown here
to coincide with the basic microscopic (diagrammatic) result, that is:

\begin{eqnarray}
&\sigma _{DOS}^{TDGL}\equiv -2n_{s}\frac{e^{2}}{m^{\ast }}\tau _{SO}\equiv 
\notag \\
&- \frac{e^{2}}{\hbar d}\frac{7\zeta \left( 3\right) }{\pi ^{4}}\int \frac{%
d\left( \xi \left( T\right) ^{2}q^{2}\right) }{\varepsilon +\xi \left(
T\right) ^{2}q^{2}}=\sigma _{xx}^{LV\left( 5+7\right) }
\end{eqnarray}

\section{The dispersion suppressed zero-field DOS conductivity}

Consider the corrected zero-field DOS conductivity, Eq.\ref{sig_DOS^LVmod} ,
which includes the $q$ dependence of the single-electron relaxation time $%
\tau _{SO}\left( q\right) $, that is: $\ $%
\begin{equation*}
\sigma _{DOS}^{LV,corr}=\frac{e^{2}}{2\pi ^{4}}\frac{1}{\hbar d}%
\int_{0}^{1}dx\psi ^{\prime \prime }\left( \frac{1}{2}+\frac{2}{\pi ^{2}}%
x\right) \frac{1}{\varepsilon +x}
\end{equation*}%
where $x=\pi \hbar Dq^{2}/8k_{B}T$, and: $\psi ^{\prime \prime }\left(
1/2+2x/\pi ^{2}\right) =-2\sum \limits_{n=0}^{\infty }\left( n+1/2+2x/\pi
^{2}\right) ^{-3}$. \ The result is nonvanishing negative function of $%
\varepsilon $ ($0<\varepsilon \ll 1$), which may be estimated, using
two-parameter fitting scheme, to be:

\begin{equation}
\sigma _{DOS}^{LV,corr}\approx -\left( \frac{1}{2\pi ^{4}}\right) \left( 
\frac{e^{2}}{d\hbar }\right) \left[ 7\ln \left( 1+\frac{1}{\varepsilon }%
\right) -1\right]  \label{sig_DOS^LV,corr}
\end{equation}

It shows that the $q$ dispersion of $\tau _{SO}\left( q\right) $ suppresses
the magnitude of $\sigma _{DOS}^{LV}$ to about $1/2$ of Eq.\ref{sig_DOS^LVq0}

\section{The CPFs density}

An expression for the CPFs density in terms of a generalized form of the
fluctuation energy function $\Phi \left( q;H\right) $, which is valid at
large wavenumbers, is given by (see Ref.\cite{MZJPC2023}): 
\begin{equation*}
n_{CPF}\left( H\right) =\left( \frac{7\zeta \left( 3\right) E_{F}}{4\pi
^{2}k_{B}T}\right) \frac{\pi }{d}\int_{0}^{q_{c}^{2}}\frac{d\left(
q^{2}\right) }{\left( 2\pi \right) ^{2}}\frac{1}{\Phi \left( q;H\right) }
\end{equation*}%
where $\Phi \left( q;H\right) $ is given by Eq.\ref{Phi(x;H)}. Note the use
of the general expression, Eq.\ref{Phi(x;H)}, for the dimensionless energy $%
\Phi \left( q;H\right) $ under the integral for $n_{S}\left( H\right) $,
with the field-independent kinetic energy variable $x$, suggesting that the
cutoff wavenumber $q_{c}$ should also be field independent.

Under the dirty limit condition: $\hbar /\tau _{SO}=\varepsilon _{SO}>>\mu
_{B}H$, one finds: 
\begin{eqnarray*}
&&f_{+}\simeq \frac{\varepsilon _{SO}}{4\pi k_{B}T}\left[ 1+\left( 1-2\left( 
\frac{\mu _{B}H}{\varepsilon _{SO}}\right) ^{2}\right) \right] \simeq \frac{%
\varepsilon _{SO}}{2\pi k_{B}T}, \\
&&f_{-}\simeq \frac{\varepsilon _{SO}}{4\pi k_{B}T}\left[ 1-\left( 1-2\left( 
\frac{\mu _{B}H}{\varepsilon _{SO}}\right) ^{2}\right) \right] \simeq \\
&&\frac{\varepsilon _{SO}}{2\pi k_{B}T}\left( \frac{\mu _{B}H}{\varepsilon
_{SO}}\right) ^{2}<<f_{+},
\end{eqnarray*}%
and for $a_{\pm }$: $a_{+}\simeq 1,a_{-}\simeq \left( \frac{\mu _{B}H}{%
\varepsilon _{SO}}\right) ^{2}<<1$, so that: {\small 
\begin{eqnarray}
&&\Phi \left( q;H\right) \simeq \Phi \left( x,\frac{T_{H}}{T}\right) \equiv
\label{Phi(x;T_H/T)} \\
&&\varepsilon _{H}+\psi \left[ \frac{1}{2}\left( 1+\frac{T_{H}}{T}\right) +x%
\right] -\psi \left[ \frac{1}{2}\left( 1+\frac{T_{H}}{T}\right) \right] 
\notag
\end{eqnarray}%
} where $T_{H}$ is given in Eq.\ref{T_H}.

The linear approximation: 
\begin{eqnarray}
\Phi \left( x,\frac{T_{H}}{T}\right) &\simeq &\Phi _{L}\left( x,\frac{T_{H}}{%
T}\right) =\varepsilon _{H}+\psi ^{\prime }\left[ \frac{1}{2}\left( 1+\frac{%
T_{H}}{T}\right) \right] x  \notag \\
&\simeq &\varepsilon _{H}+\eta \left( H\right) x  \label{Phi_L(x;T_H/T)}
\end{eqnarray}%
is identical to the energy denominator of the fluctuation propagator in Eq.%
\ref{D(q)^H}.

The exact expression for the CPFs density is, then, written in the form:

\begin{equation}
n_{CPF}\left( H\right) =n_{CPF}^{0}\int_{0}^{x_{c0}T_{c0}/T}\frac{dx}{\Phi
\left( x,\frac{T_{H}}{T}\right) }  \label{n_S(H)exact}
\end{equation}%
where $n_{S0}$ is given in Eq.\ref{n_s0}.

Performing the integration in Eq.\ref{n_S(H)exact} with $\Phi \left(
x,T_{H}/T\right) $ replaced by the linear approximation $\Phi _{L}\left(
x,T_{H}/T\right) $ we find for the linear approximation of the CPF density:

\begin{equation}
n_{CPF}^{L}\left( H\right) =\frac{n_{CPF}^{0}}{\psi ^{\prime }\left[ \frac{1%
}{2}\left( 1+\frac{T_{H}}{T}\right) \right] }\ln \left( 1+\frac{x_{c}\left(
H\right) }{\varepsilon _{H}}\right)  \label{n_S^L}
\end{equation}%
where: 
\begin{equation*}
x_{c}\left( H\right) \equiv \psi ^{\prime }\left[ \frac{1}{2}\left( 1+\frac{%
T_{H}}{T}\right) \right] \left( \frac{T_{c0}}{T}\right) x_{c0}
\end{equation*}

Using the asymptotic form of the digamma function for the low temperature
limit $T/T_{H}\ll 1$ we find:

\begin{eqnarray}
n_{CPF}^{L}\left( H\right) &\rightarrow &\left( \frac{T_{H}}{2T}\right)
n_{CPF}^{0}\ln \left( 1+\frac{x_{c}\left( H\right) }{\varepsilon _{H}}%
\right) ,  \label{n_S^LT0} \\
x_{c}\left( H\right) &\rightarrow &\left( \frac{T_{c0}}{T_{H}}\right) x_{c0}
\notag
\end{eqnarray}%
which is identical to Eq.\ref{n_S(H)T0}, provided $x_{c}\left( H\right) $ is
selected to be field independent, e.g.: $x_{c}\left( H\right) \rightarrow 1$%
. The latter selection is consistent with the selection of the cutoff
wavenumber to depend on field according to: $q_{c}=\xi ^{-1}\left( H\right) $
(see Eq.\ref{xi(H)} and text around Eq.\ref{sig_DOS(H)^mic}).

We note that in our fitting procedure, employed in Refs.\cite{MZPRB2021}, 
\cite{MZJPC2023}, we have used the field-independent cutoff selection for $%
q_{c}$. Specifically, the best fitting value was found to agree with $%
x_{c0}=0.015$ (see Eq.\ref{xc0}) with the selected values of the other
parameters: $T_{c0}=0.212\boldsymbol{\mathit{K}}$ and $\varepsilon
_{SO}=3\times 10^{-3}\boldsymbol{\mathit{eV}}$.

As indicated in the main text, deviations from the linear approximation may
be significant for large cutoff wavenumbers which do not satisfy condition %
\ref{linapprox}. Under these circumstances and in the low temperatures
region, $T/T_{H}\ll 1$, we may use the asymptotic form of the digamma
functions in Eq.\ref{Phi(x;T_H/T)}, which can be rewritten as: $\Phi \left(
x,\frac{T_{H}}{T}\right) \rightarrow \varepsilon _{H}+\ln \left(
x+T_{H}/2T\right) -\ln \left( T_{H}/2T\right) =\varepsilon _{H}+\ln \left(
1+2xT/T_{H}\right) $, that is:

\begin{eqnarray}
\Phi \left( x,\frac{T_{H}}{T}\right) &\rightarrow &\varepsilon _{H}+\ln
\left( 1+\frac{\hbar Dq^{2}}{2\pi k_{B}T_{H}}\right) ,  \label{PhiAsymp} \\
T/T_{H} &\ll &1  \notag
\end{eqnarray}

In Fig.3 we plot $n_{CPF}\left( H\right) $ using Eq.\ref{n_S(H)exact} with
the exact expression, Eq.\ref{Phi(x;T_H/T)}, for $\Phi \left(
x,T_{H}/T\right) $, together with $n_{CPF}^{L}\left( H\right) $ given in Eq.%
\ref{n_S^L}, as functions of $T_{c0}/T$ for different values of the field $H$
and cutoff parameter $x_{c0}$.

\section{The TDGL AL conductivity}

The AL conductivity is calculated within the TDGL functional approach from
the time-ordered current-current correlator given in Eq.\ref{Q_ALgen}, which
can be rewritten in the form: 
\begin{eqnarray}
&Q_{AL}\left( i\Omega _{\nu }\right) =k_{B}T\left( \frac{2e}{\hbar }\right)
^{2}\left( \frac{1}{2\pi d}\right) \int \limits_{0}^{x_{c}}xdx
\label{Q^TDGL_AL} \\
&\times \sum \limits_{k=0,\pm 1,\pm 2,....}\frac{\Phi ^{\prime }\left(
x+\left \vert k+\nu \right \vert /2\right) }{\Phi \left( x+\left \vert k+\nu
\right \vert /2\right) }\frac{\Phi ^{\prime }\left( x+\left \vert k\right
\vert /2\right) }{\Phi \left( x+\left \vert k\right \vert /2\right) }  \notag
\end{eqnarray}

This is done by the analytic continuation from the imaginary Matsubara
frequency $i\Omega _{\nu }$ to the real frequency $\omega $ , i.e.: $%
Q_{AL}\left( i\Omega _{\nu }\right) \rightarrow Q_{AL}^{R}\left( \omega
\right) $, so that in the static limit: $\sigma _{AL}^{TDGL}=\lim_{\omega
\rightarrow 0}\left( i/\omega \right) \left[ Q_{AL}^{R}\left( \omega \right)
-Q_{AL}^{R}\left( 0\right) \right] $. \ 

It is interesting to note that under direct analytic continuation of the
discrete summation in Eq.\ref{Q^TDGL_AL} about zero frequency, i.e. $\nu
\rightarrow \hbar \omega /2\pi ik_{B}T\rightarrow 0$, all nonzero
Matsubara-frequency terms are cancelled out and the remaining $k=0$ term can
be written in the form: 
\begin{equation}
\sigma _{AL}^{TDGL}=\left( \frac{e^{2}}{8\pi ^{2}\hbar d}\right) \int
\limits_{0}^{x_{c}}\left( \frac{\Phi ^{\prime }\left( x\right) }{\Phi \left(
x\right) }\right) ^{2}dx  \label{sig_ALd}
\end{equation}

Exploiting the linear approximation, i.e.: \ $\Phi \left( x\right) \simeq
\varepsilon _{H}+\eta \left( H\right) x$, and performing the integration
over $x$ we find:%
\begin{equation}
\sigma _{AL}^{TDGL}\simeq \left( \frac{e^{2}}{16\hbar d}\right) \frac{%
\widetilde{\eta }\left( H\right) }{\varepsilon _{H}\left( 1+\frac{%
\varepsilon _{H}}{\eta \left( H\right) x_{c}}\right) }  \label{sig_ALlin}
\end{equation}%
in full agreement with Eq.\ref{sig_AL(H)^mic}. Note the factor of $1/2$
multiplying the integer variables $\left \vert k+\nu \right \vert $ and $%
\left \vert k\right \vert $ in Eq.\ref{Q^TDGL_AL}, which was erroneously
missing in a similar expression for this correlator in Refs.\cite{MZPRB2021},%
\cite{MZJPC2023} and consequently leading to an error by a factor of $2$ in
the calculation of the AL conductivity there.

\section{CPFs localization length}

Starting with Eq.\ref{g(rho)} for the correlation function $g\left( \rho
/\xi _{0}\right) $ we will be interested here in its asymptotic behavior for 
$\rho /\xi _{0}\gg 1$, which will enable us to clearly identify the
parameters that determine the localization of CPFs. Exploiting the linear
approximation: 
\begin{equation}
\Phi \left( q;H\right) \simeq \widetilde{\varepsilon }_{H}+\frac{2}{\pi ^{2}}%
\psi ^{\prime }\left( 1/2+\frac{T_{H}}{T}\right) \frac{T_{c0}}{T}\xi
_{0}^{2}q^{2}  \label{Phi_L(q;H)}
\end{equation}%
and denoting:\ $\zeta \equiv \widetilde{q}/\sqrt{t},t\equiv T_{c0}/T$, we
have: 
\begin{equation}
g\left( \rho /\xi _{0}\right) \simeq \int_{0}^{q_{c}\xi _{0}/\sqrt{t}}\frac{%
\zeta J_{0}\left[ \zeta \left( \sqrt{t}\rho /\xi _{0}\right) \right] }{%
\widetilde{\varepsilon }_{H}+\psi ^{\prime }\left( \frac{1}{2}+\frac{T_{H}}{T%
}\right) \frac{2}{\pi ^{2}}\zeta ^{2}}d\zeta  \label{g_L(rho)}
\end{equation}

Using the asymptotic form of the Bessel function, $J_{0}\left( z\right) \sim
\left( 2\pi z\right) ^{-1/2}e^{\pm iz\mp i\pi /4}$ under the integral in Eq.%
\ref{g_L(rho)} and focusing only on the pole contributions at:

\begin{equation*}
\zeta =\pm i\zeta _{pole}=\pm i\sqrt{\frac{\pi ^{2}}{2}\frac{\widetilde{%
\varepsilon }_{H}}{\psi ^{\prime }\left( \frac{1}{2}+\frac{T_{H}}{T}\right) }%
}
\end{equation*}%
we may estimate the asymptotic behavior:

\begin{equation}
g\left( \rho /\xi _{0}\right) \sim A\frac{\exp \left[ -\left( \rho /\xi
_{0}\right) \sqrt{\frac{\pi ^{2}}{2}\frac{T}{T_{c0}}\frac{\widetilde{%
\varepsilon }_{H}}{\psi ^{\prime }\left( \frac{1}{2}+\frac{T_{H}}{T}\right) }%
}\right] }{\left( \frac{T}{T_{c0}}\widetilde{\varepsilon }_{H}\right) ^{1/4}%
\left[ \psi ^{\prime }\left( \frac{1}{2}+\frac{T_{H}}{T}\right) \right]
^{3/4}}  \label{g_asymp}
\end{equation}%
where $A$ is an adjustable parameter of the order one (see Fig.4).

It is dominated by the decaying exponential with the characteristic length:

\begin{equation*}
\rho _{loc}\left( H\right) =\xi _{0}/\sqrt{\frac{\pi ^{2}}{2}\frac{T}{T_{c0}}%
\frac{\widetilde{\varepsilon }_{H}}{\psi ^{\prime }\left( \frac{1}{2}+\frac{%
T_{H}}{T}\right) }}
\end{equation*}%
which is identical to Eq.\ref{rho_loc} in the main text. Its well-defined
low-temperature limit can be easily found from the asymptotic form of the
digamma function, $\psi ^{\prime }\left( 1/2+T_{H}/T\right) \sim T/T_{H}$ ,
that is:%
\begin{eqnarray*}
\rho _{loc}\left( H\right) &\rightarrow &\left[ \frac{2}{\pi ^{2}}\left( 
\frac{T_{c0}}{T_{H}}\right) \frac{1}{\widetilde{\varepsilon }_{H}}\right]
^{1/2}\xi _{0}; \\
T &\ll &T_{H}
\end{eqnarray*}

\end{document}